\documentclass[twocolumn]{autart}

\usepackage{amsmath}
\usepackage{graphicx}
\usepackage{amssymb}
\usepackage{verbatim}
\usepackage{epsfig}
\usepackage{subfig}
\usepackage{enumitem}
\usepackage{enumerate}
\usepackage{cite}
\usepackage{epstopdf}

\newtheorem{rmrk}{Remark}
\newtheorem{assm}{Assumption}
\newtheorem{lemma}{Lemma}

\def\rea{\mathbb{R}}

\def\calc{\mathcal{C}}

\def\calb{\mathcal{B}}

\def\cald{\mathcal{D}}

\def\call{\mathcal{L}}
\def\calr{\mathcal{R}}

\def\hal{{1 \over 2}}

\def\begequarr{\begin{eqnarray}}
\def\endequarr{\end{eqnarray}}
\def\begequarrs{\begin{eqnarray*}}
\def\endequarrs{\end{eqnarray*}}
\def\begarr{\begin{array}}
\def\endarr{\end{array}}
\def\begequ{\begin{equation}}
\def\endequ{\end{equation}}
\def\lab{\label}
\def\begdes{\begin{description}}
\def\enddes{\end{description}}
\def\begenu{\begin{enumerate}}
\def\begite{\begin{itemize}}
\def\endite{\end{itemize}}
\def\endenu{\end{enumerate}}

\def\lef[{\left[\begin{array}}
\def\rig]{\end{array}\right]}

\def\begcen{\begin{center}}
\def\endcen{\end{center}}
\def\begrem{\begin{rmrk}\rm}
\def\endrem{\end{rmrk}}

\def\l2{{\mathcal L}_2}
\def\l2e{{\cal L}_{2e}}

\def\rea{\mathbb{R}}

\def\col{\mbox{col}}

\def\lef[{\left[\begin{array}}
\def\rig]{\end{array}\right]}

\def\TAC{{\it IEEE Trans. Automatic Control}}

\def\SCL{{\it Systems and Control Letters}}

\def\CST{{\it IEEE Trans. Control Systems Technology}}

\def\begmat#1{\begin{bmatrix}#1\end{bmatrix}}
\def\begali#1{\begin{align}{#1}\end{align}}
\def\begalis#1{\begin{align*}{#1}\end{align*}}

\usepackage{color}

\usepackage[prependcaption,colorinlistoftodos]{todonotes}

\graphicspath{{fig/}}

\begin{document}

\begin{frontmatter}

\title{Permanent Magnet Synchronous Motors are Globally Asymptotically Stabilizable with PI Current Control\thanksref{footnoteinfo}}

\thanks[footnoteinfo]{This paper was not presented at any IFAC 
meeting. Corresponding author R. Ortega.}

\author[SPLC]{Romeo Ortega}\ead{ortega@lss.supelec.fr},
\author[UG2]{Nima Monshizadeh}\ead{n.monshizadeh@rug.nl},
\author[UG]{Pooya Monshizadeh}\ead{p.monshizadeh@rug.nl},  \qquad
\author[ITMO]{Dmitry Bazylev}\ead{bazylev@corp.ifmo.ru},
\author[ITMO]{Anton Pyrkin}\ead{pyrkin@corp.ifmo.ru}
\address[SPLC]{Laboratoire des Signaux et Systèmes, CNRS-SUPELEC, Plateau du Moulon, 91192, Gif-sur-Yvette, France}
\address[UG2]{Engineering and Technology Institute, University of Groningen, 9747AG, The Netherlands.}
\address[UG]{Bernoulli Institute for Mathematics and Computer Science, University of Groningen, 9747AG, The Netherlands}
\address[ITMO]{Department of Control Systems and Informatics, ITMO University, Kronverksky av., 49, 197101, Saint Petersburg, Russia}


\maketitle

\begin{abstract}
This note shows that the industry standard desired equilibrium for permanent magnet synchronous motors ({\em i.e.}, maximum torque per Ampere) can be globally asymptotically stabilized with a PI control around the current errors, provided some viscous friction (possibly small) is present in the rotor dynamics and the proportional gain of the PI is suitably chosen. Instrumental to establish this surprising result is the proof that the map from voltages to currents of the incremental model of the motor satisfies some passivity properties. The analysis relies on basic Lyapunov theory making the result available to a wide audience. 
\end{abstract}

\end{frontmatter}

\section{Introduction}
\lab{sec1}
%
Control of electric motors is achieved in the vast majority of commercial drives via nested loop PI controllers \cite{LEO,KRA,PAR}: the inner one wrapped around current errors and an external one that defines the desired values for these currents to generate a desired torque---for speed or position control. The rationale to justify this control configuration relies on the, often reasonable, assumption of time-scale separation between the electrical and the mechanical dynamics. In spite of its enormous success, to the best of our knowledge, a rigorous theoretical analysis of the stability of this scheme has not been reported. The main contribution of this paper is to (partially) fill-up this gap for the widely popular permanent magnet synchronous motors (PMSM), proving that the inner loop PI controller ensures global asymptotic stability (GAS) of the closed-loop, provided some viscous friction (possibly arbitrarily small) is present in the rotor dynamics, that the load torque is known and the proportional gain of the PI is suitably chosen, {\em i.e.}, sufficiently high. The assumption of known load torque is later relaxed proposing an adaptive scheme that, in the spirit of the aforementioned outer-loop PI, generates, via the addition of a simple integrator, an estimate for it---preserving GAS of the new scheme. 

Several globally stable position and velocity controllers for PMSMs have been reported in the control literature---even in the sensorless context, {\em e.g.}, \cite{BODetal,LEEetal,TOMVER,TOMVERtac} and references therein. However, these controllers have received an, at best, lukewarm reception within the electric drives community, which overwhelmingly prefers the aforementioned nested-loop PI configuration. Several versions of PI schemes based on fuzzy control, sliding modes or neural network control have been intensively studied in applications journals, see \cite{JUNetal} for a recent review of this literature. To the best of our knowledge, a rigorous stability analysis of all these schemes is conspicuous by its absence. 

The importance of disposing of a complete theoretical analysis in engineering practice can hardly be overestimated. Indeed, it gives the user additional confidence in the design and provides useful guidelines in the difficult task of commissioning the controller. The interest of our contribution is further enhanced by the fact that the analysis relies on basic Lyapunov theory, using the natural (quadratic in the increments) Lyapunov function. Various attempts to establish such a result for PMSMs have been reported in the literature either relying on linear approximations of the motor dynamics or including additional terms that cancel some nonlinear terms, see \cite{HERetalijc,HERetalcst} and references therein---a standing assumption being, similarly to us, the existence of viscous friction. 

The remainder of this paper is organised as follows. The models of the PMSM are given in Section \ref{sec2}. The problem formulation is introduced in Section \ref{sec3}. The passivity of the PMSMs incremental model and the PI controller are established in Section \ref{sec4}. The main stability results are  provided in Section \ref{sec5}. Some simulation results are presented in Section \ref{sec6}. Finally, some concluding remarks and discussion of future research are given in Section~\ref{sec7}. 

\noindent {\bf Notation.}  For $x \in \rea^n$, $A \in \rea^{n \times n}$, $A>0$ we denote $|x|^2=x^\top  x$,  $\|x\|_A^2:=x^\top  A x$. For the distinguished vector $x^\star \in \mathbb{R}^n$ and a mapping $\calc: \mathbb{R}^n \to \rea^{n \times n}$, we define the constant  matrix $\calc^\star:=\calc(x^\star)$.
%
%
\section{Motor Models}
\lab{sec2}
%
In this section we present the motor model, define the desired equilibrium and give its incremental model.
\subsection{Standard $dq$ model}
\lab{subsec21}
The dynamics of the surface-mounted PMSM in the $dq$ frame is described by \cite{KRA,PETetal}:
\begin{equation}
\label{sys}
	\begin{split}
		L_d {d i_d \over dt}
		&=
		-R_si_d + \omega L_qi_q + v_d \\
		L_q {d i_q \over dt}
		&=
		-R_si_q - \omega L_di_d - \omega\Phi + v_q \\
		J{d \omega \over dt}
		&= -R_m \omega+
		n_p
		\left[
		(L_d-L_q)i_di_q + \Phi i_q
		\right] - \tau_L
	\end{split}
\end{equation}
where $i_d, i_q$ are currents, $v_d,v_q$ are voltage inputs, $\omega$ is the electrical angular velocity\footnote{Related with the rotor speed $\omega_m$ via $\omega={2n_p \over 3}\omega_m$}, ${2n_p \over 3}$ is the number of pole pairs, $L_d>0, L_q>0$ are the stator inductances, $\Phi>0$ is the back emf constant, $R_s>0$ is the stator resistance, $R_m>0$ is the viscous friction coefficient, $J>0$ is the moment of inertia and $\tau_L$ is a constant load torque.

Defining the state and control vectors as
$$
x:=\begmat{i_d \\ i_q \\ \omega},\; u:=\begmat{v_d \\ v_q},
$$ 
the system \eqref{sys} can be written in compact form as
$$
\cald \dot x + [\calc(x)+\calr]x = G u + d,
$$
where
\begalis{
 \cald &:=\begmat{L_d & 0 & 0 \\ 0 & L_q & 0  \\ 0  & 0 & {J \over n_p}}>0, \calr :=\begmat{R_s & 0 & 0 \\ 0 & R_s & 0\\ 0 & 0 & {R_m \over n_p}}>0\\ 
 \calc(x)&:=\begmat{0 & 0 & - L_q x_2 \\ 0 & 0 & L_d x_1+\Phi\\ L_q x_2 & - (L_d x_1 + \Phi) & 0}=-\calc^\top (x)\\ 
G &:=\begmat{1 & 0 \\ 0 & 1 \\ 0  & 0},\;d := \begmat{0 \\ 0 \\ -{\tau_L \over n_p}},
}

Besides simplifying the notation, the interest of the representation above is that it reveals the power balance equation of the system. Indeed, the total energy of the motor is
$$
H(x)=\hal x^\top  \cald x,
$$
whose derivative yields
\begequ
\lab{powbal}
\underbrace{\dot H}_{stored\;power} = -\underbrace{x^\top  \calr x}_{dissipated} + \underbrace{y^\top  u}_{supplied} - \underbrace{x_3 {\tau_L \over n_p},}_{extracted}
\endequ
where we used the skew-symmetry of $\calc(x)$ and defined the currents as outputs, that is, 
$$
y:=G^\top  x=\begmat{i_d \\ i_q}.
$$
The current-feedback PI design is analysed in this paper viewing it as a passivity-based controller (PBC)---a term that was coined in \cite{ORTSPO}---where the main idea is to preserve a power balance equation like the one above but now with a new stored energy and a new dissipation term. This objective is accomplished in two steps, the shaping of the systems energy to give it a desired form, {\em i.e.}, to have a minimum at the desired equilibrium, and the injection of damping. The shaped energy function qualifies, then, as a Lyapunov function that ensures stability of the equilibrium, which can be rendered asymptotically stable via the damping injection. 

\begrem
\lab{rem1}
See \cite{ORTetal,VAN} for additional discussion on the general theory of PBC and its practical applications and \cite{ARAetal,ZHAetal} for some recent developments on PID-PBC.
\endrem
\subsection{Incremental model}
\lab{subsec22}
The industry standard desired equilibrium is the maximum torque per Ampere value defined as
\begequ
\lab{equ}
x^\star :=\col\left(0,{1 \over n_p \Phi}(\tau_L+R_m \omega^\star ),\omega^\star \right),
\endequ
where $\omega^\star $ is the desired electrical speed. With respect to this equilibrium we define the incremental model
\begali{
\nonumber
\cald \dot {\tilde x} + \calc(x) \tilde x +[\calc(x) -\calc^\star  ]x^\star   + \calr \tilde x & = G \tilde u\\
\tilde y &=G^\top \tilde x, 
\lab{incsys}
}
where $\tilde {(\cdot)}:= {(\cdot)} - {(\cdot)}^\star $, {$\calc^\star:=\calc(x^\star)$}, and we used the fact that
\begalis{
(\calc^\star + \calr) x^\star   &= G u^\star  +d\\
y^\star   &=G^\top x^\star,
}
with 
\[
u^\star = \begmat{-{1 \over n_p \Phi}L_q\omega^\star (\tau_L+R_m \omega^\star ) \\ \Phi \omega^\star +{1 \over n_p \Phi}R_s(\tau_L+R_m \omega^\star )}.
\]
Note that
\begequ
\lab{tily}
\tilde y=\begmat{x_1 \\ x_2-{1 \over n_p \Phi}(\tau_L+R_m \omega^\star )}.
\endequ
%
%
\section{Problem Formulation}
\lab{sec3}
%
We are interested in the paper in giving conditions for GAS of a PI controller wrapped around the currents $i_d,i_q$, which are assumed to be measurable. We consider two different scenarios. 

\begdes
\item[S1] Known $\tau_L$, $\Phi$ and $R_m$ and ``classical" PI
\begali{
\nonumber
\dot x_c &= \tilde y \\
\lab{pi}
u &= -K_I x_c - K_P \tilde y,
}
with $\tilde y$ defined in \eqref{tily} and $K_I,K_P>0$.\\

\item[S2] Unknown $\tau_L$ but verifying the following (reasonable) assumption.

\begin{assm}\em
\lab{ass1}
A positive constant $\tau_L^{\tt max}$ such that
$$
|\tau_L|  \leq \tau_L^{\tt max},
$$
is known.
\end{assm}

Moreover, we assume that $\omega$ is measurable and, besides knowing the parameters $\Phi$ and $R_m$, it is also assumed that $L_d,L_q$ and $J$ are known.\footnote{As shown in Proposition \ref{pro2}, these additional assumptions are needed to design the estimator of $\tau_L$.} In this scenario, we consider the adaptive PI controller
\begali{
\nonumber
\dot x_c &= \begmat{x_1 \\ x_2-\hat x_2^\star } \\
\lab{pice}
u &= -K_I x_c - K_P \begmat{x_1 \\ x_2-\hat x_2^\star },
}
with $K_I,K_P>0$, where $\hat x_2^\star $ is an estimate of the reference $q$-current $x_2^\star $, generated from an estimator of the simple integral form
\begali{
\nonumber
\dot \chi&=f(x,\chi)\\
\lab{obsdes}
\hat x_2^\star &= h(x,\chi),
}
with $\chi \in \rea$, which is to be designed.
\enddes

In both scenarios we want to prove that there exists a positive-definite gain matrix $K_P^{\tt min}$ such that the PMSM model \eqref{sys} in closed-loop with the PI controller \eqref{pi} or \eqref{pice} with  $K_P \geq K_P^{\tt min}$ has a GAS equilibrium at  $(x^\star,x_c^\star,\chi^\star )$ for some  $x_c^\star  \in \rea^2$ and $\chi^\star  \in \rea$ such that $h(x^\star ,\chi^\star )= x_2^\star$. Moreover, in the second scenario,  $K_P^{\tt min}$ should not depend on $\tau_L$, but only on the bound given in Assumption \ref{ass1}.

\begrem
\lab{rem5}
As indicated in the introduction, in practice the reference value for $x_2$ is generated with an outer-loop PI around speed errors, that is,
\begali{
\nonumber
\dot \chi &= \tilde x_3 \\
\lab{nespi}
\hat x_2^\star &= -a_I \chi  - a_P \tilde x_3,
}
with $a_I,a_P>0$. Unfortunately, the stability analysis of this configuration is far from obvious and we will need to propose another form for the functions $f(x,\chi)$ and  $h(x,\chi)$ in \eqref{obsdes}.
\endrem

\begrem
\lab{remrm}
For the sake of completeness we also propose an estimator for the viscous friction coefficient $R_m$, which generates a consistent estimate under an excitation assumption. See Subsection \ref{subsec53}. 
\endrem
%
\section{Passivity Analysis}
\lab{sec4}
%
%
\subsection{Dissipativity of the incremental model}
\lab{subsec41}
In this section we give conditions under which the incremental model \eqref{incsys} satisfies a dissipation inequality of the form
\begin{equation}
\label{disine}
\dot {U} \leq \epsilon |\tilde y|^2 + \tilde y^\top  \tilde u.
\end{equation}
with
\begequ
\lab{tilh}
U(\tilde x)=\hal \|\tilde x\|^2_\cald.
\endequ
for some $\epsilon \in \rea$. If $\epsilon$ is negative it is then said that the incremental model of the system \eqref{sys} is output strictly passive, if it is positive, then it is called output feedback passive, indicating the shortage of passivity \cite{JAYetal,MONetal,VAN}.\footnote{In \cite{MONetal,VAN} the property of passivity of the incremental model is called shifted passivity.} 

Comparing \eqref{disine} with the open-loop power balance equation \eqref{powbal} we see that, besides removing the term of extracted power, we have shaped the energy---assigning a minimum at the desired equilibrium $x^\star  $---and replaced the damping term $x^\top  \calr x$ by $\epsilon |\tilde y|^2$. Notice that, if $\epsilon$ is positive, it is easy to add damping selecting a control $\tilde u = -K_P \tilde y$, with $K_P=k_p I_2>0.$ Indeed, this yields a damping term $-(k_p - \epsilon)|\tilde y|^2$, with $k_p>\epsilon$ we ensure $\dot {U} \leq 0$---whence, stability of the equilibrium.  As explained in Remark \ref{rem4}, a more clever option is to add an integral action, yielding a PI.

\begin{lemma}\em
\lab{lem1}
Define the matrix
$$
\calb:=\begmat{2R_s+2\epsilon & {(L_d-L_q)}x_3^\star   &  {-}L_d x_2^\star    \\ {(L_d-L_q)}x_3^\star   & 2R_s+2\epsilon & 0\\  {-}L_d x_2^\star   & 0 & 2 {R_m \over n_p}}, 
$$
for some $\epsilon \in \rea$. If $\calb \geq 0$ the dissipation inequality \eqref{disine} holds. 
\end{lemma}
\begin{pf}
Computing the derivative of \eqref{tilh} along the solutions of \eqref{incsys} we get
\begalis{
\dot {U} &= -\tilde x^\top [\calc(x)-\calc^\star  ]x^\star  -\tilde x^\top  \calr \tilde x + \tilde y^\top  \tilde u\\ &= -\hal \tilde x^\top  (\calb-2\epsilon GG^\top ) \tilde x + \tilde y^\top  \tilde u\\ &= -\hal \tilde x^\top  \calb \tilde x +\epsilon |\tilde y|^2 + \tilde y^\top  \tilde u,
}
where we have used the fact that
$$
[\calc(x)-\calc^\star  ]x^\star  =\begmat{0 & -L_q x_3^\star   & 0  \\ L_d x_3^\star   & 0 & 0\\ - L_d x_2^\star   & 0 & 0}\tilde x,
$$
to get the second identity and use the definition of $\tilde y$ given in \eqref{incsys} in the third identity. The proof is completed imposing the condition $\calb \geq 0$.
\end{pf}

\begrem
\lab{rem2}
Lemma 1 follows as a direct application of Proposition 1 and Remark 3 of \cite{MONetal}, where passivity of the incremental model of general port-Hamiltonian systems with strictly convex energy function is studied. To make the present paper self-contained we have included a proof of the lemma.
\endrem

\begrem
\lab{rem3}
A dissipativity analysis similar to Lemma \ref{lem1}  has been carried out within the context of transient stability of power systems in \cite{ORTnolcos}, for synchronous generators connected to a constant voltage source in \cite{VAN2} and \cite{CALetal}. In all these papers the shifted Hamiltonian of \cite{JAYetal}, which in these cases boils down to the natural incremental energy function, is also used to establish stability conditions---that involve the analysis of positivity of a damping matrix similar to $\calb$.  
\endrem

\subsection{Strict passivity of the PI controller}
\lab{subsec42}
%
In this subsection we prove the input strict passivity of the PI controller. Although this result is very well-known \cite{VAN,ZHAetal}, a proof is given here for the sake of completeness.

\begin{lemma}\em
\lab{lem2}
Given any constant $y_c^\star  \in \rea^2$, define the error signal $ \tilde y_c:=y_c-y_c^\star$. The PI controller
\begalis{
\dot x_c &= u_c \\
y_c &= K_I x_c + K_P u_c,
}
defines an {input strictly passive} map $u_c \mapsto  \tilde y_c$, with storage function 
\begequ
\lab{hc}
H_c(\tilde x_c):=\hal \|\tilde x_c\|^2_{K_I},
\endequ
where $x^\star _c:=-K_I^{-1} y_c^\star.$ More precisely
$$
\dot H_c = -\|u_c\|^2_{K_P} + u_c^\top \tilde y_c.
$$ 
\end{lemma}

\begin{pf}
First, notice that, using the definition of $x_c^\star$ in $y_c$ we have that
\begequ
\lab{tilu}
\tilde y_c = K_I \tilde x_c + K_P u_c.
\endequ
Computing the derivative of $H_c$ along the trajectories of \eqref{pi} yields
\begalis{
\dot H_c&=\tilde x_c^\top  K_I u_c = u_c^\top (\tilde y_c -K_P u_c),
}
where we have used \eqref{tilu} in the second identity, which completes the proof.
\end{pf}

\begrem
\lab{rem4}
The PI controller described above will be coupled with the PMSM via the (power-preserving) interconnection $u_c=\tilde y$ and $y_c=-u$. Lemma \ref{lem2} shows the interest of adding an integral action: there is no need to know $u^\star$ to implement the controller. 
\endrem
%
\section{Main Results}
\lab{sec5}

%
\subsection{Stability of the standard PI controller}
\lab{subsec51}
%
\begin{prop}\em
\lab{pro1}
Consider the PMSM model \eqref{sys} in closed-loop with the PI controller \eqref{pi}, the integral gain $K_I>0$ and the proportional gain\footnote{$K_P$ is taken of this particular form to simplify the presentation of the main result---this choice is done without loss of generality.} $K_P=k_p I_2>0.$ There exists a positive constant $k_p^{\tt min}$ such that
\begequ
\lab{kpmin}
k_p \geq k_p^{\tt min}
\endequ
ensures that $(x^\star,x_c^\star )$ is a GAS equilibrium of the closed-loop system. For non-salient PMSM, {\em i.e.}, when $L_d=L_q$, the constant $k_p^{\tt min}$ can be chosen such that
\begequ
\lab{kp-nonsalient}
k_p^{\tt min}>\,{L_d^2\over 4 R_m n_p \Phi^2}\left(\tau_L+R_m|\omega^\star |\right)^2- {R_s}
\endequ
\end{prop}

\begin{pf}
Summing up \eqref{tilh} and \eqref{hc} define the positive definite, radially unbounded, Lyapunov function candidate
\begequ
\lab{w}
W(\tilde x,\tilde x_c):=U(\tilde x)+H_c(\tilde x_c).
\endequ
Computing its derivative we get
\begali{
\lab{dotw}
\dot W &= -\hal \|\tilde x\|^2_\calb +(\epsilon-k_p) |\tilde y|^2 = -{\frac{1}{2}}\|\tilde x\|^2_{\calr_d},
}
where we defined the matrix
$$
\calr_d:= \begmat{2R_s+2k_p & {(L_d-L_q)}\omega^\star   & - L_d x_2^\star    \\ {(L_d-L_q)}\omega^\star   & 2R_s+2k_p & 0\\ - L_d x_2^\star   & 0 & 2 {R_m \over n_p}}
$$
From \eqref{dotw} we immediately conclude that if $\calr_d>0$, then the equilibrium $(x^\star ,x_c^\star )$ is globally stable. Moreover, invoking Krasovskii's Theorem, we prove that the equilibrium is GAS because
$$
\tilde x(t) \equiv 0 \Rightarrow \tilde x_c(t) \equiv 0.
$$
The gist of the proof is then to prove the existence of the lower bound $k_p^{\tt min}$ that ensures positivity of $\calr_d$.

Towards this end, we recall the following well-known (Schur complement) equivalence:
$$
\begmat{A & B\\ B^\top & C} >0 \quad \Leftrightarrow \quad C > 0 \;\mbox{and}\; A-BC^{-1}B^\top > 0.
$$
Directly applying this to $\calr_d$ with
$$
A:=\begmat{2R_s+2k_p & {(L_d-L_q)}\omega^\star   \\ {(L_d-L_q)}\omega^\star   & 2R_s+2k_p },\; B:=\begmat{ - L_d x_2^\star \\ 0 },
$$
and $C:=2 {R_m \over n_p}$, shows that $\calr_d>0$ if and only if
\begequ
\lab{sch}
(R_s+k_p)I_2 >\hal\begmat{{n_pL^2_d |x_2^\star|^2 \over 2 R_m } & {(L_q-L_d)}\omega^\star     \\ {(L_q-L_d)}\omega^\star   & 0}.
\endequ
This proves the existence of $k_p^{\tt min}$ such that, if \eqref{kpmin} holds then $\calr_d>0$. In case $L_d=L_q$, $k_p^{\tt min}$ can be chosen as in \eqref{kp-nonsalient}.
\end{pf}
\subsection{An asymptotically stable adaptive PI controller}
\lab{subsec52}
%
In applications the load torque $\tau_L$, and consequently $x_2^\star $ are unknown. It is, therefore, necessary to replace its value above by an estimate, a task, that is accomplished in the proposition below.

\begin{prop}\em
\lab{pro2}
Consider the PMSM model \eqref{sys} verifying Assumption \ref{ass1} in closed-loop with the adaptive PI controller \eqref{pice} with the estimator
\begali{
\nonumber
J\dot {\chi} &= -R_m \omega+n_p\left[(L_d-L_q)i_di_q + \Phi i_q\right] -\ell(\chi-\omega)\\
\nonumber
{\hat \tau}_L &=  \ell(\chi-\omega)\\
\hat x_2^\star&={1 \over n_p \Phi}(\hat \tau_L+R_m \omega^\star )
\lab{tauobs}
}
where $\ell>0$. Fix the proportional gain as $K_P=k_p I_2>0.$ 

There exists a positive constant $k_p^{\tt min}$---dependent only on $\tau_L^{\tt max}$---such that
\eqref{kpmin} ensures that $(x^\star,x_c^\star,\chi^\star )$, with $\chi^\star :={\tau_L \over \ell}+\omega^\star $ is a GAS equilibrium of the closed-loop system. 
\end{prop}

\begin{pf}
Similarly to the proof of Proposition \ref{pro1}, we first need to prove that $\calr_d>0$. This follows immediately invoking \eqref{sch} and noting that, from the definition of the equilibria in \eqref{equ}, we have
$$
{1 \over n_P \Phi}(|\tau_L| + R_m|\omega^\star |) \geq |x_2^\star |.
$$
Thus, a $k_p^{\tt min}$ that depends only on $\tau_L^{\tt max}$, can readily be defined.

We proceed now to prove that the estimator \eqref{tauobs} generates an exponentially convergent estimate of $\tau_L$.  Defining the estimation error $e_{\tau_L}:= \hat \tau_L - \tau_L,$ the error dynamics yields
\begequ
\lab{dynetau}
\dot e_{\tau_L} = -{\ell \over J} e_{\tau_L},
\endequ
which is clearly exponentially stable for all $\ell>0$. 

To simplify the presentation of the analysis of  the overall error dynamics let us define the reference output error signal
$$
e_{y^\star }:=\hat y^\star  - y^\star =\begmat{ 0 \\ \hat x_2^\star  - x_2^\star }={1 \over n_P\Phi}\begmat{ 0 \\ e_{\tau_L} },
$$
which replaced in \eqref{pice} yields
\begalis{
\dot {\tilde x}_c & =  \tilde y - e_{y^\star } \\
\tilde u &= -K_I \tilde x_c - K_P(\tilde y - e_{y^\star })
}
The closed-loop is then a cascaded dynamics of the form
\begali{
\nonumber
\dot e_{y^*} & = -{\ell \over n_P \Phi J} e_{y^*}\\
\dot \xi &= f(\xi)+\begmat{\cald^{-1}GK_P\\-I_2} e_{y^\star }
\lab{cassys}
}
with $\xi:=\col(\tilde x,\tilde x_c)$ and the dynamics $\dot \xi = f(\xi)$ has the origin as a GAS equilibrium.

The GAS proof is completed invoking Theorem 1 of \cite{PANLOR} that shows that the cascaded system is globally stable, which implies that all trajectories are bounded. GAS follows immediately from the well-known fact \cite{SEISUA} that the cascade of two GAS systems is GAS if all trajectories are bounded.\footnote{The first author expresses his gratitude to Antoine Chaillet, Denis Efimov and Elena Panteley for several discussions on the topic of cascaded systems.}
\end{pf}
\subsection{A globally convergent estimator of $R_m$}
\lab{subsec53}
%
In the lemma below we show that it is possible to add an adaptation term to estimate the friction coefficient $R_m$, that is usually uncertain, provided some excitation conditions are satisfied. 

\begin{lemma}\em
\lab{lem3}
Consider the mechanical equation in \eqref{sys} and the gradient estimator
\begequ
\lab{estrm}
\dot {\hat R}_m=\gamma \phi(z - \hat R_m \phi),
\endequ
with $\gamma>0$ an adaptation gain and the measurable signals
\begali{
\nonumber
z&:={\beta p^2 \over (p+\alpha)^2}[J\omega]+{\beta p \over (p+\alpha)^2}[n_p(L_q-L_d)i_di_q-n_p\Phi i_q]\\
\lab{yphi}
\phi&:={\beta p \over (p+\alpha)^2}[\omega],
}
where $p:={d \over dt}$ and $\alpha, \beta>0$. The following equivalence holds true
$$
\phi \notin \call_2\;\Leftrightarrow\;\lim_{t \to \infty}|\hat R_m(t)-R_m|=0,
$$
with $\call_2$ the space of square integrable functions.
\end{lemma}

\begin{pf}
Applying the filter ${\beta p \over (p+\alpha)^2}$ to the mechanical equation in \eqref{sys}, recalling that $\tau_L$ is constant, and using the definitions \eqref{yphi} yields the linear regression model 
$$
z = R_m \phi + \epsilon_t
$$
where $\epsilon_t$ is an exponentially decaying term stemming form the filters initial conditions, which can be neglected without loss of generality. Replacing the equation above in \eqref{estrm} yields the error equation
\begequ
\lab{errequ}
\dot e_{R_m}=-\gamma \phi^2 e_{R_m},
\endequ
where $e_{R_m};=\hat R_m - R_m$ is the parameter estimation error. The proof is completed integrating \eqref{errequ}.
\end{pf}

\begrem
\lab{rem7}
As always in estimation problems some kind of excitation on the signals must be imposed to guarantee convergence. In our case it is the condition of non-square integrability of $\omega$, which is weaker than the more classical persistence of excitation assumption---in which case the convergence of the parameter error is exponential. 
\endrem

\begrem
\lab{rem8}
An alternative to the estimators presented above is to add a nonlinear integral action to compensate for both unknowns $\tau_L$ and $R_m$ as done in \cite{FERetal}. In any case, both options considerably complicate the control law, a scenario that is beyond the scope of this paper. Also, although it is possible to carry out the stability analysis of the combination of the estimators of $\tau_L$ and $R_m$, we avoid this discussion for the aforementioned reason.
\endrem

%
%
\section{Simulation Results}
\begin{figure}
\begin{center}
\includegraphics[width=7cm]{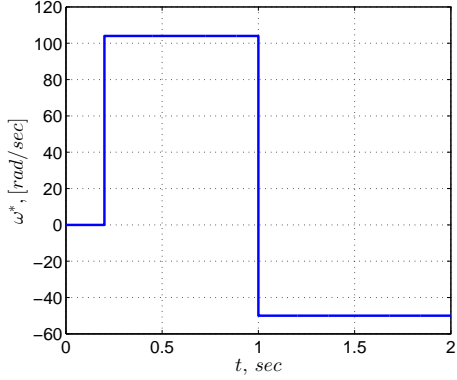}    
\caption{Speed reference} 
\label{fig_1}
\end{center}
\end{figure}

\begin{figure}
\begin{center}
\includegraphics[width=7cm]{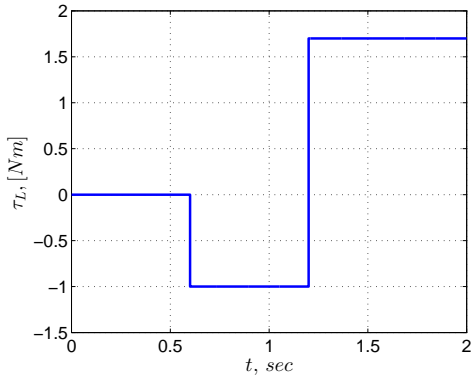}    
\caption{Torque load input} 
\label{fig_2}
\end{center}
\end{figure}

\lab{sec6}
%
The objective of the simulations is to verify numerically the performance of the proposed controllers under different gains and external signals. First, we consider a constant speed reference and load input, then, to illustrate the tracking ability of the load torque estimator, we propose the time-varying profiles depicted in Figs \ref{fig_1} and \ref{fig_2}. In all cases we took the PI gains as $K_P=k_p I_2$ and  $K_I=k_i I_2$.


The following scenarios were considered.
\begite
\item[(C1)] Inner-loop PI \eqref{pi} with known $\tau_L$ and $R_m$, considering the cases of $k_p \geq k_p^{\tt min}$ and $k_p < k_p^{\tt min}$ and constant speed reference and load torque.
\item[(C2)] Adaptive inner-loop PI \eqref{pice} with load torque estimator \eqref{tauobs} for the load torque of Fig. \ref{fig_2} and the speed reference of Fig. \ref{fig_1}.
\item[(C3)] Adaptive inner-loop PI \eqref{pice} with load torque \eqref{tauobs} and viscous friction coefficient \eqref{estrm} estimators for the time-varying profiles of Figs. \ref{fig_1} and \ref{fig_2}. 
\item[(C4)] Inner-loop PI \eqref{pi} with outer-loop PI in speed error \eqref{nespi} for the time-varying profiles of Figs. \ref{fig_1} and \ref{fig_2}.
\endite
In simulations we use the motor data provided by \cite{LIU2015} with $R_m=0.02$ Nms. The motor parameters are given in Table \ref{tab_1}. 

\begin{table}[hb]
\begin{center}
\begin{tabular}{lc}
\hline
Parameter (units) &  Value \\
\hline
Rated current (A) & 4 \\
Nominal electrical speed $\omega_n$ (r/sec) & 104.7 \\
Number of pole pairs $n_p$ (--) & 3 \\ 
Direct-axis inductance $L_d$ (mH) & 31.2 \\ 
Quadrature-axis inductance $L_q$ (mH) & 55.0 \\ 
Stator resistance $R_s$ ($\Omega$) & 6 \\ 
Drive inertia $J$ (kgm$^2$) &  $3.61\times 10^{-4}$ \\ 
Permanent magnet flux $\Phi$ (Wb) &  0.236 \\ 
\hline
\end{tabular}
\vspace{2mm}
\caption{Motor data}
\lab{tab_1}
\end{center}
\end{table}

The maximum torque load of Assumption \ref{ass1} is chosen 70$\%$ higher than the rated value, which corresponds to $\tau_L^{\tt max}=4.6$ Nm. It should be noted that, for the nominal electrical speed $\omega_n=104.72$ rad/sec and this conservative value of $\tau_L^{\tt max}$, the minimal proportional gain which provides $\calr_d>0$ is $k_p^{\tt min}=-2.32$.  Implying that the incremental model of the motor is passive---that is, $\epsilon$ in Lemma \ref{lem1} is negative---and, consequently, it can be stably regulated setting $u=u^\star $. Obviously, for robustness reasons, a closed-loop PI is preferred. 

\begin{figure*}[htp]
\centering
\subfloat[][Speed error]{{\label{fig_3a}}\includegraphics[width=0.31\textwidth]{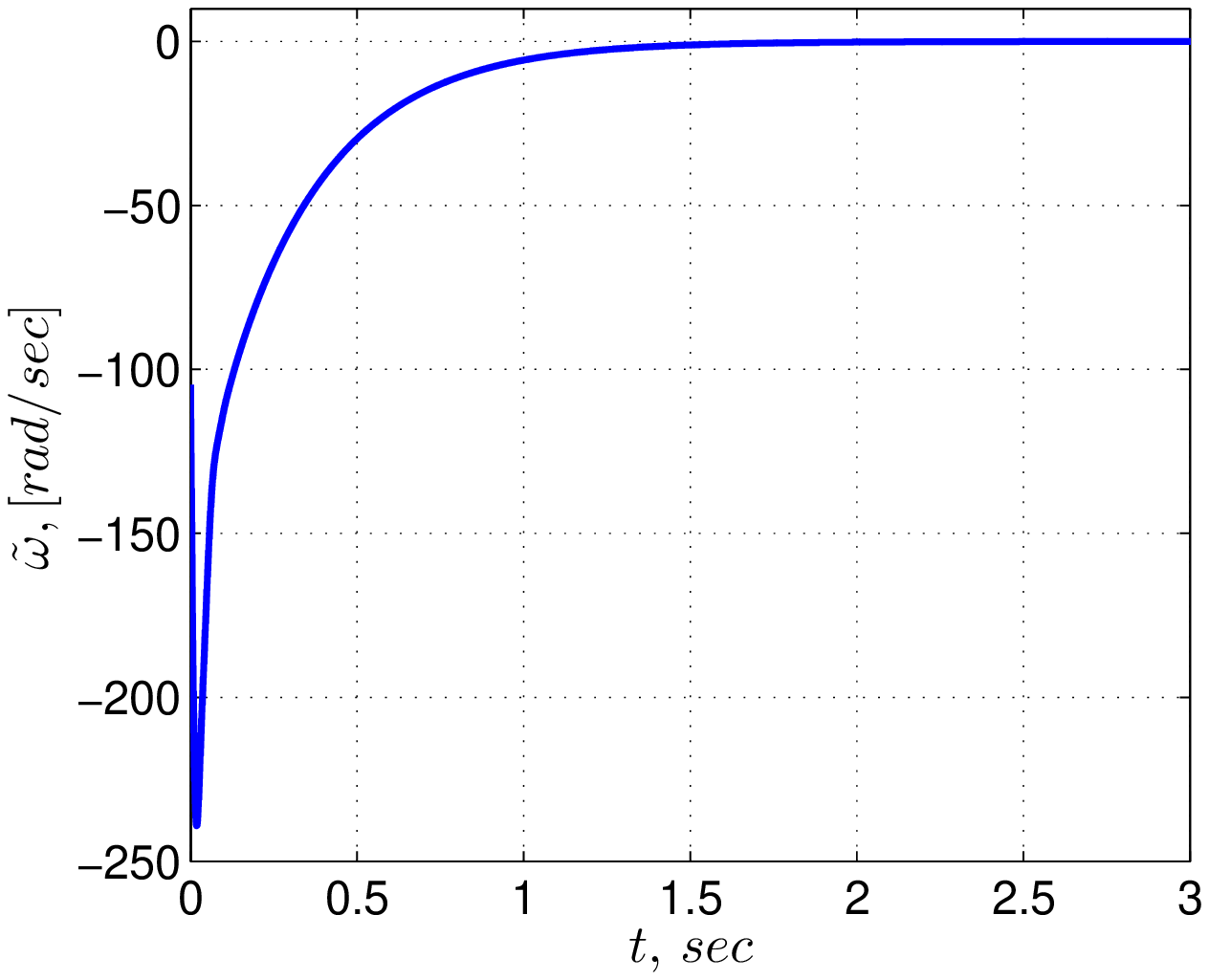}}
\
\subfloat[][Current errors $\tilde{i}_d$ (I) and $\tilde{i}_q$ (II)]{{\label{fig_3b}}\includegraphics[width=0.31\textwidth]{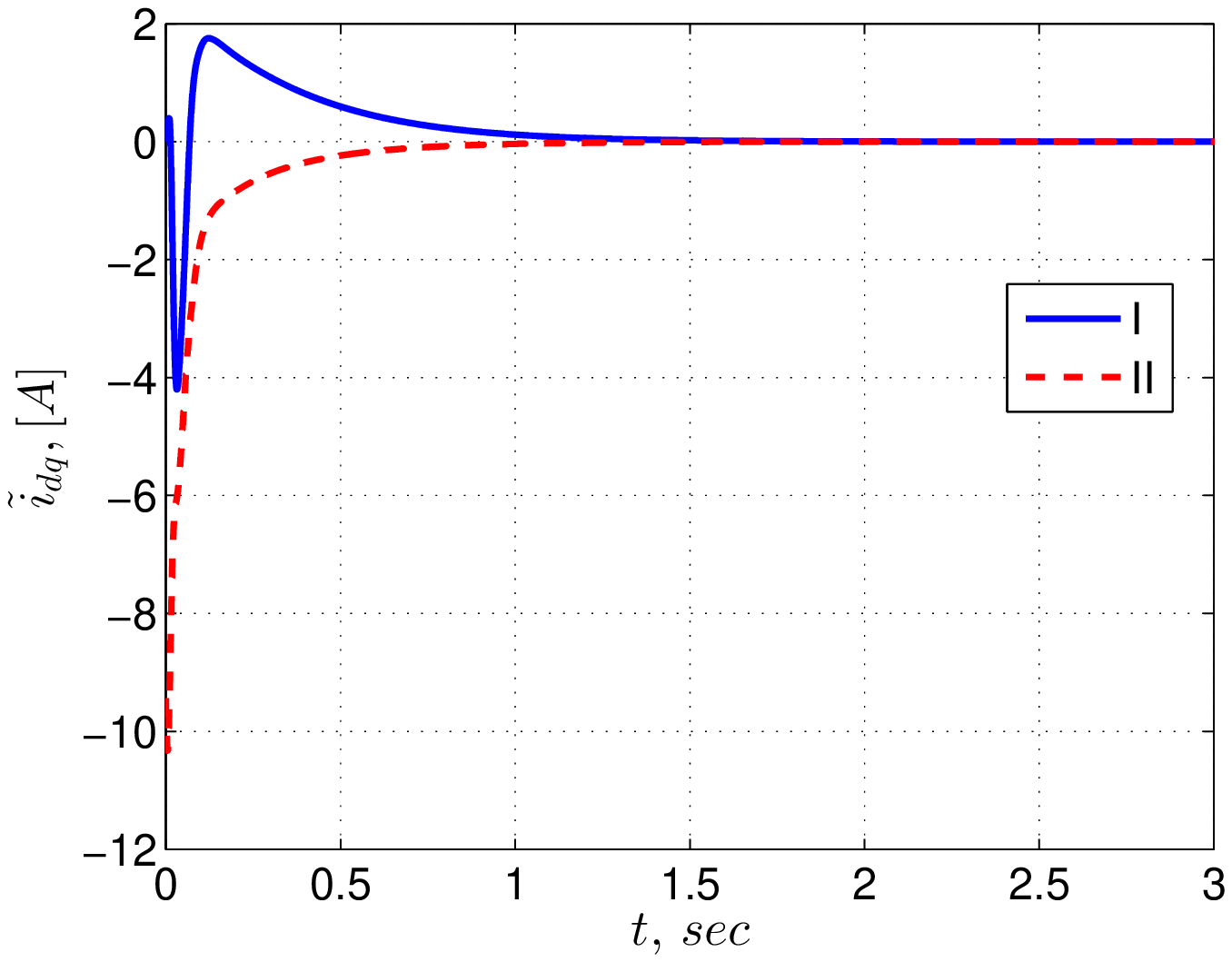}}
\
\subfloat[][Voltages $v_d$ (I) and $v_q$ (II)]{{\label{fig_3c}}\includegraphics[width=0.31\textwidth]{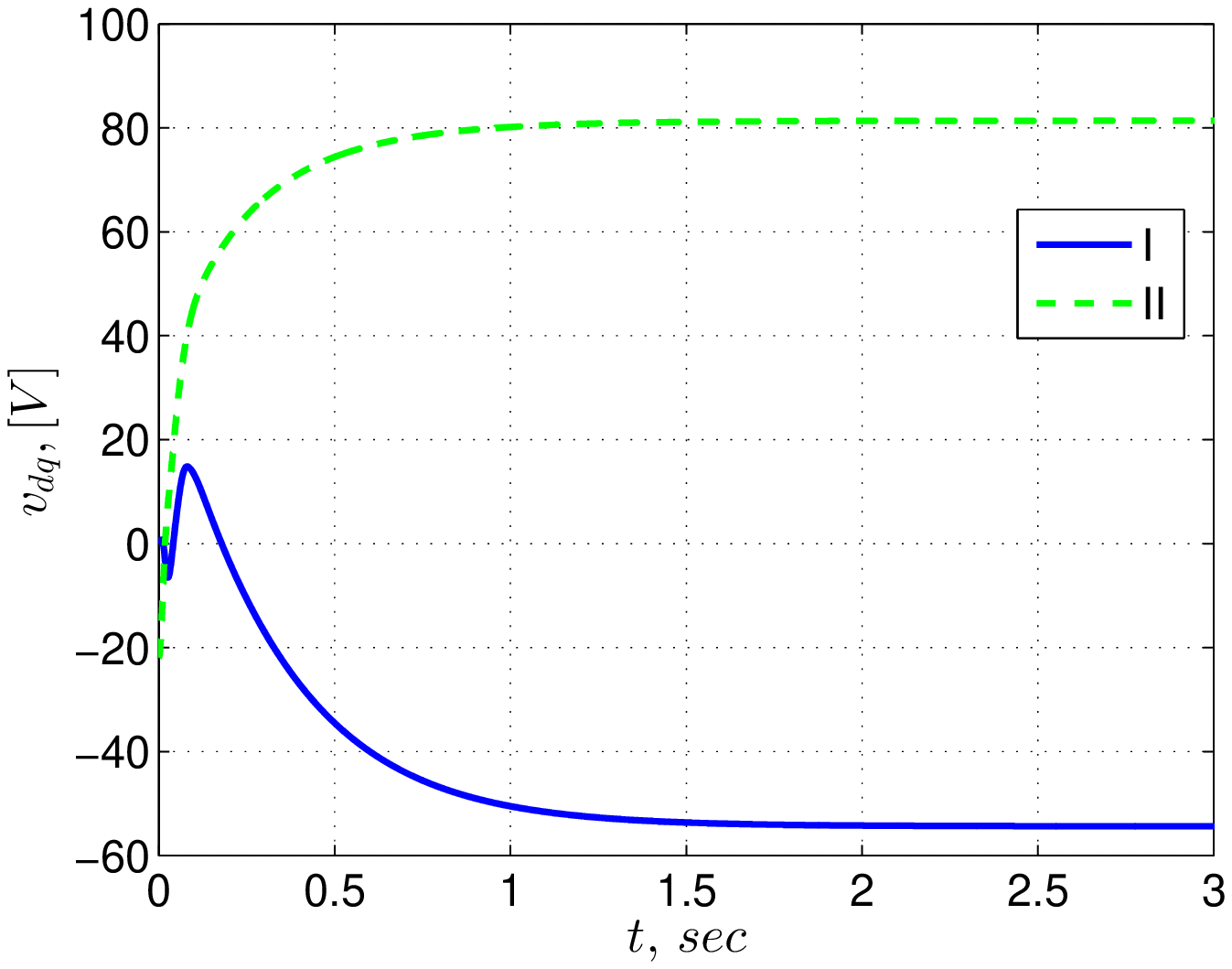}}
\vspace{-2mm}
\caption{Transients in the system with inner-loop current PI \eqref{pi}: $k_p=k_p^{\tt min}$ and $k_i=100$}
\label{fig_3}
\end{figure*}

\begin{figure*}[htp]
\centering
\subfloat[][Speed error]{{\label{fig_4a}}\includegraphics[width=0.31\textwidth]{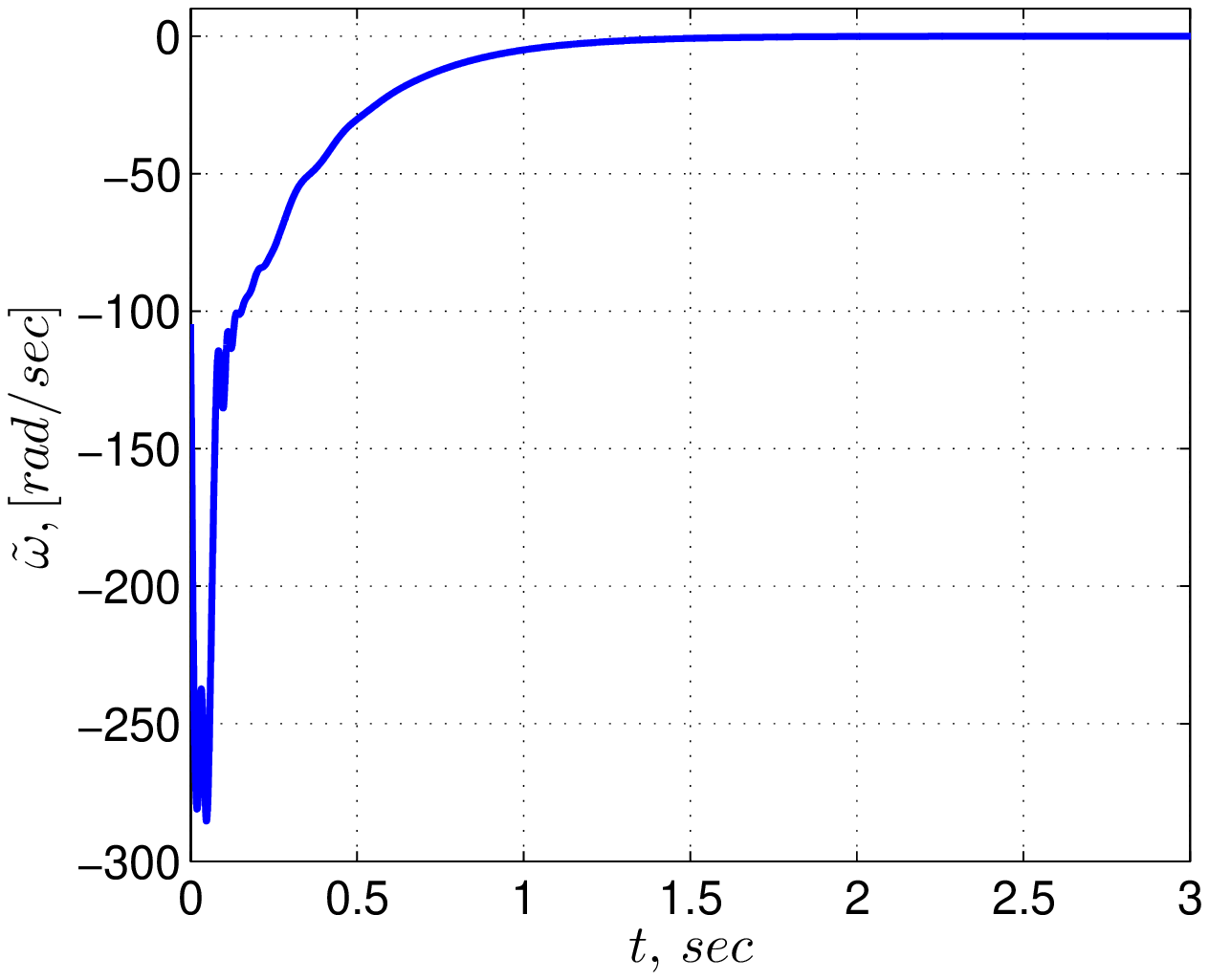}}
\
\subfloat[][Current errors $\tilde{i}_d$ (I) and $\tilde{i}_q$ (II)]{{\label{fig_4b}}\includegraphics[width=0.31\textwidth]{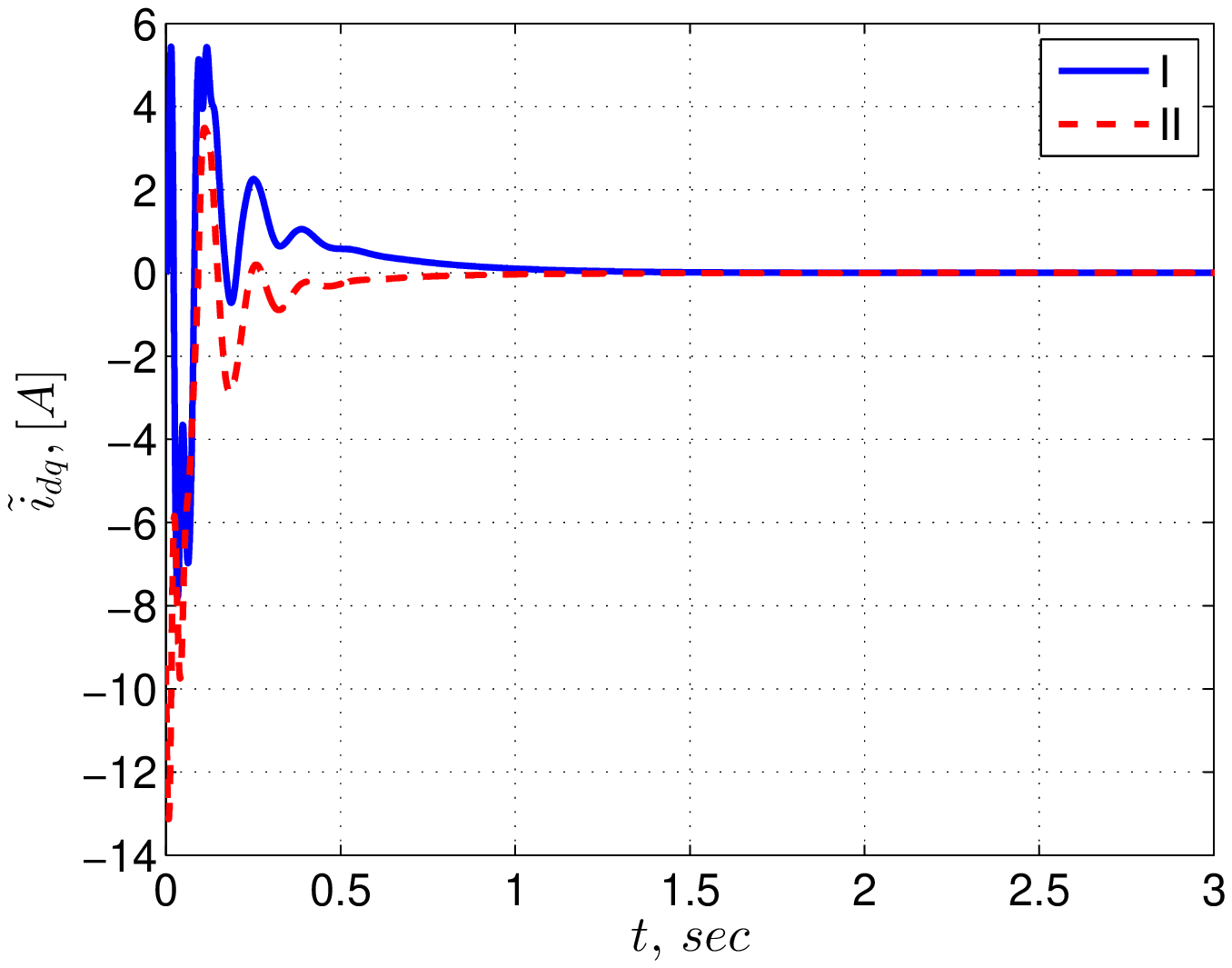}}
\
\subfloat[][Voltages $v_d$ (I) and $v_q$ (II)]{{\label{fig_4c}}\includegraphics[width=0.31\textwidth]{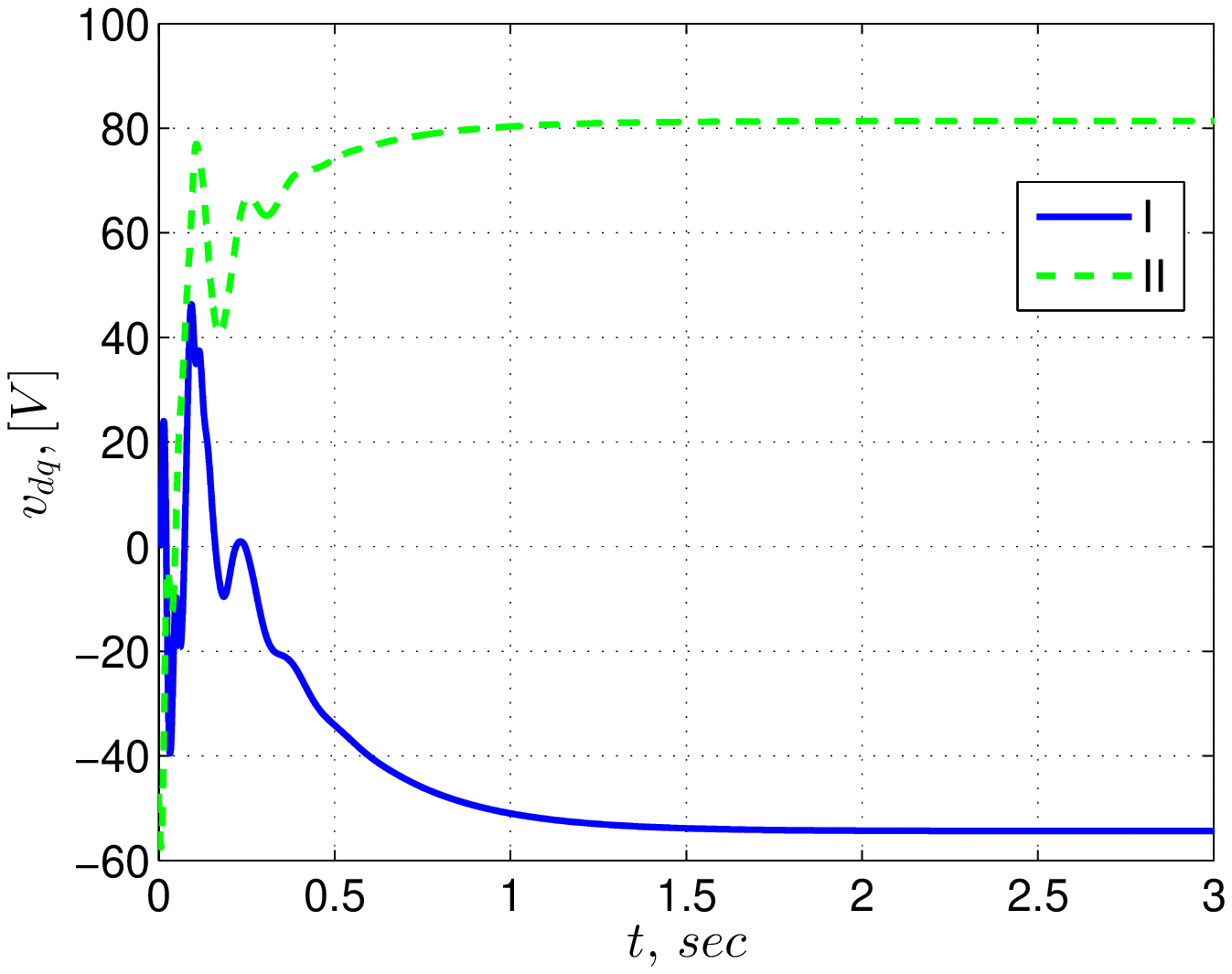}}
\vspace{-2mm}
\caption{Transients in the system with  inner-loop current PI \eqref{pi}: $k_p=-5<k_p^{\tt min}$ and $k_i=100$}
\label{fig_4}
\end{figure*}

\begin{figure*}[htp]
\centering
\subfloat[][Speed error]{{\label{fig_5a}}\includegraphics[width=0.31\textwidth]{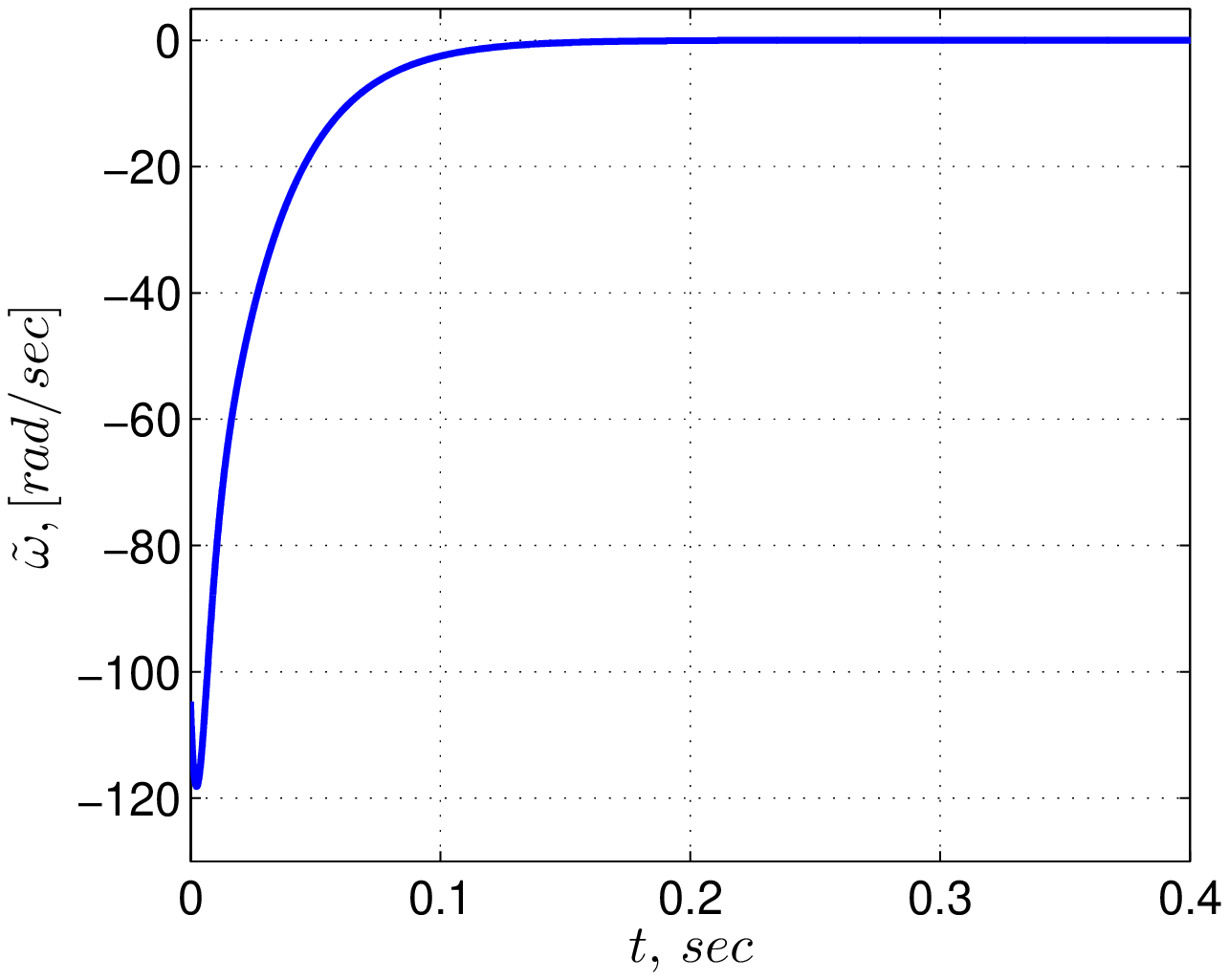}}
\
\subfloat[][Current errors $\tilde{i}_d$ (I) and $\tilde{i}_q$ (II)]{{\label{fig_5b}}\includegraphics[width=0.31\textwidth]{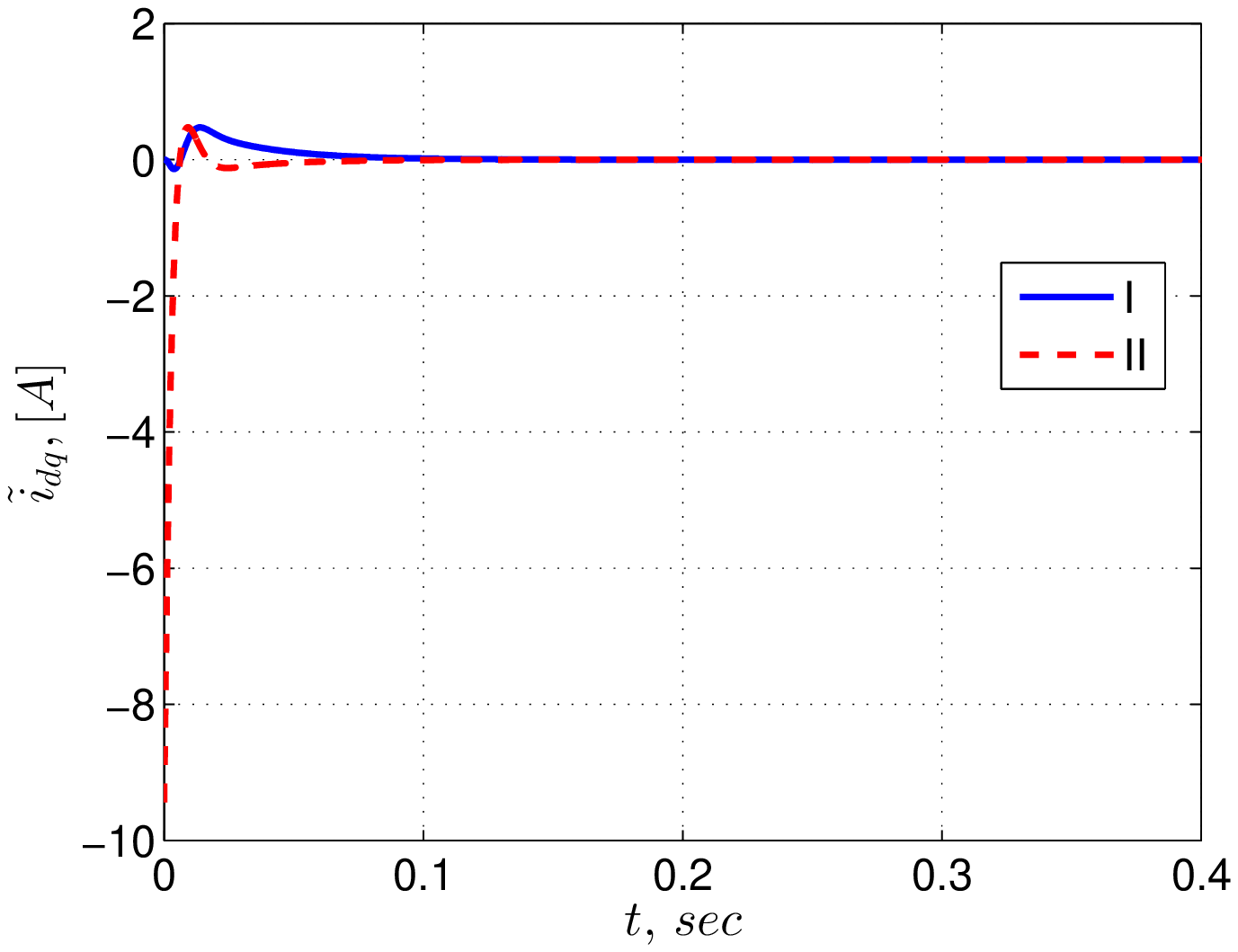}}
\
\subfloat[][Voltages $v_d$ (I) and $v_q$ (II)]{{\label{fig_5c}}\includegraphics[width=0.31\textwidth]{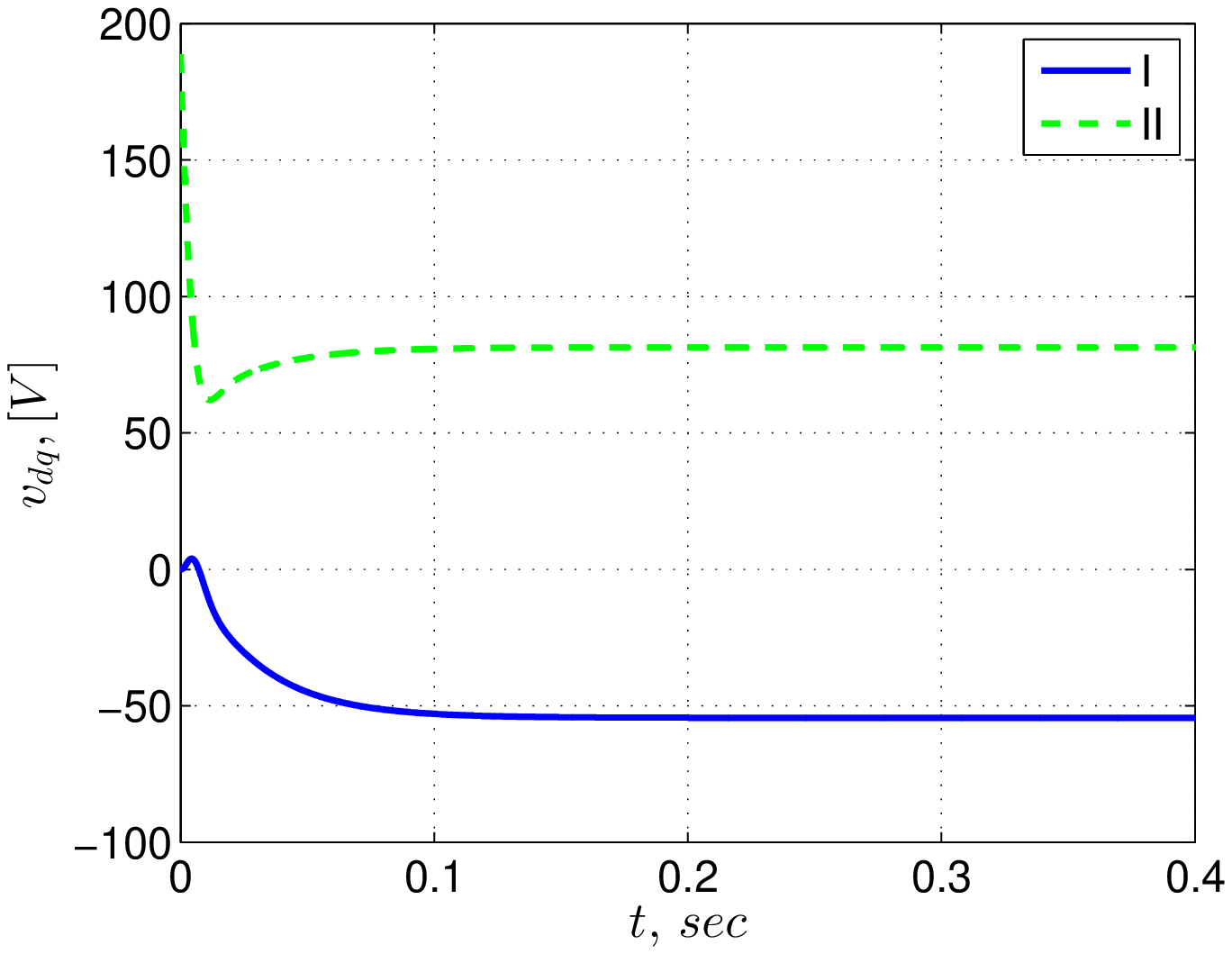}}
\vspace{-2mm}
\caption{Transients in the system with  inner-loop current PI \eqref{pi}: $k_p=20>k_p^{\tt min}$ and $k_i=4000$}
\label{fig_5}
\end{figure*}

Figs. \ref{fig_3} and \ref{fig_4} show the effect of increasing the gain $k_p$ that, as expected, improves the convergence rate. Although of no practical interest, the simulation with negative $k_p$ is presented to corroborate the theoretical result. In this respect, numerical simulations show that the motor becomes unstable for  $k_p < -5.8$. In  Fig. \ref{fig_5} both PI gains are increased obtaining a much faster response---notice the difference in time scales. In all simulations the difference in time-scales between the electrical and the mechanical dynamics is clearly apparent.

\begin{figure*}[htp]
\centering
\subfloat[][Speed error]{{\label{fig_6a}}\includegraphics[width=0.31\textwidth]{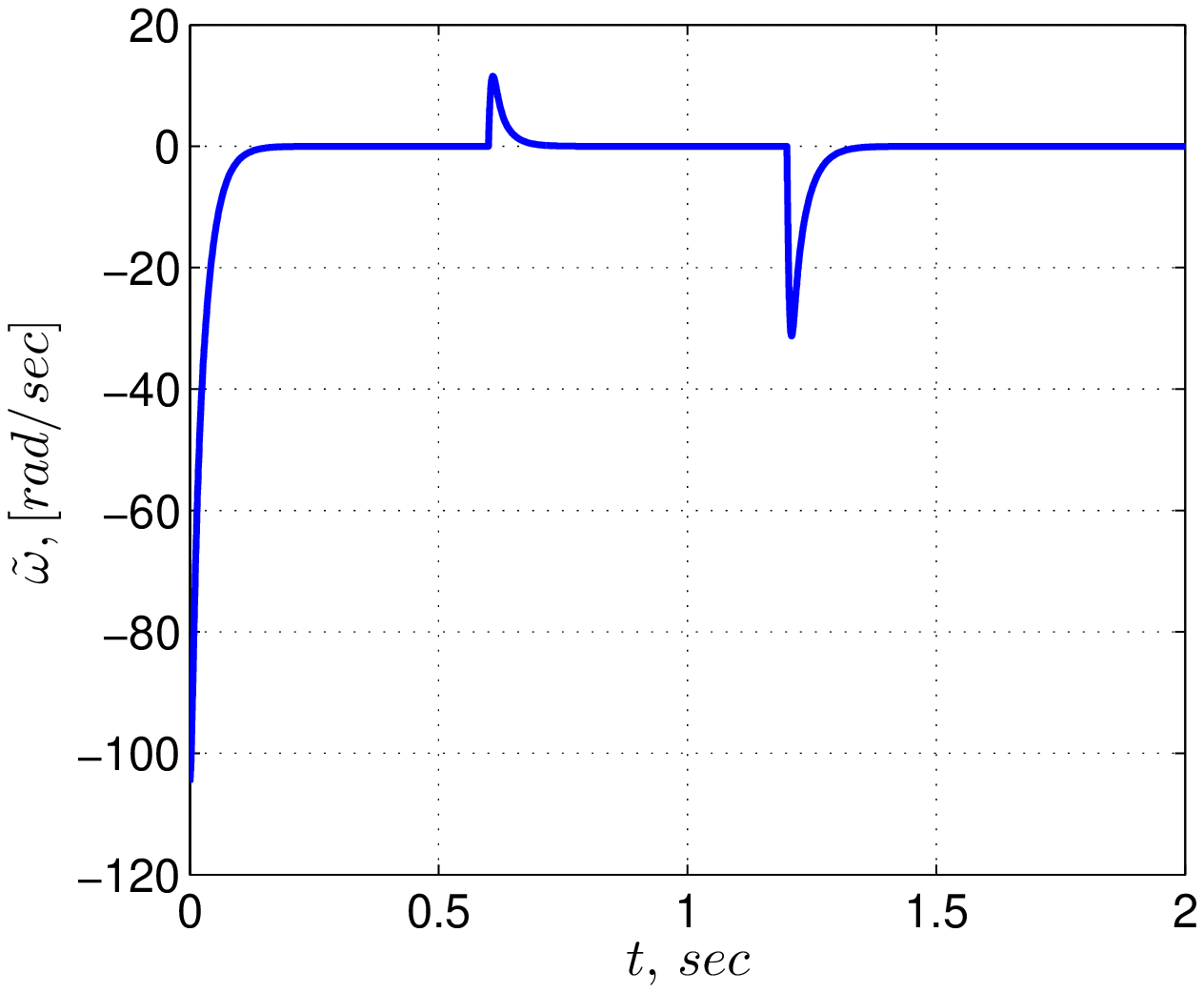}}
\
\subfloat[][Current errors $\tilde{i}_d$ (I) and $\tilde{i}_q$ (II)]{{\label{fig_6b}}\includegraphics[width=0.31\textwidth]{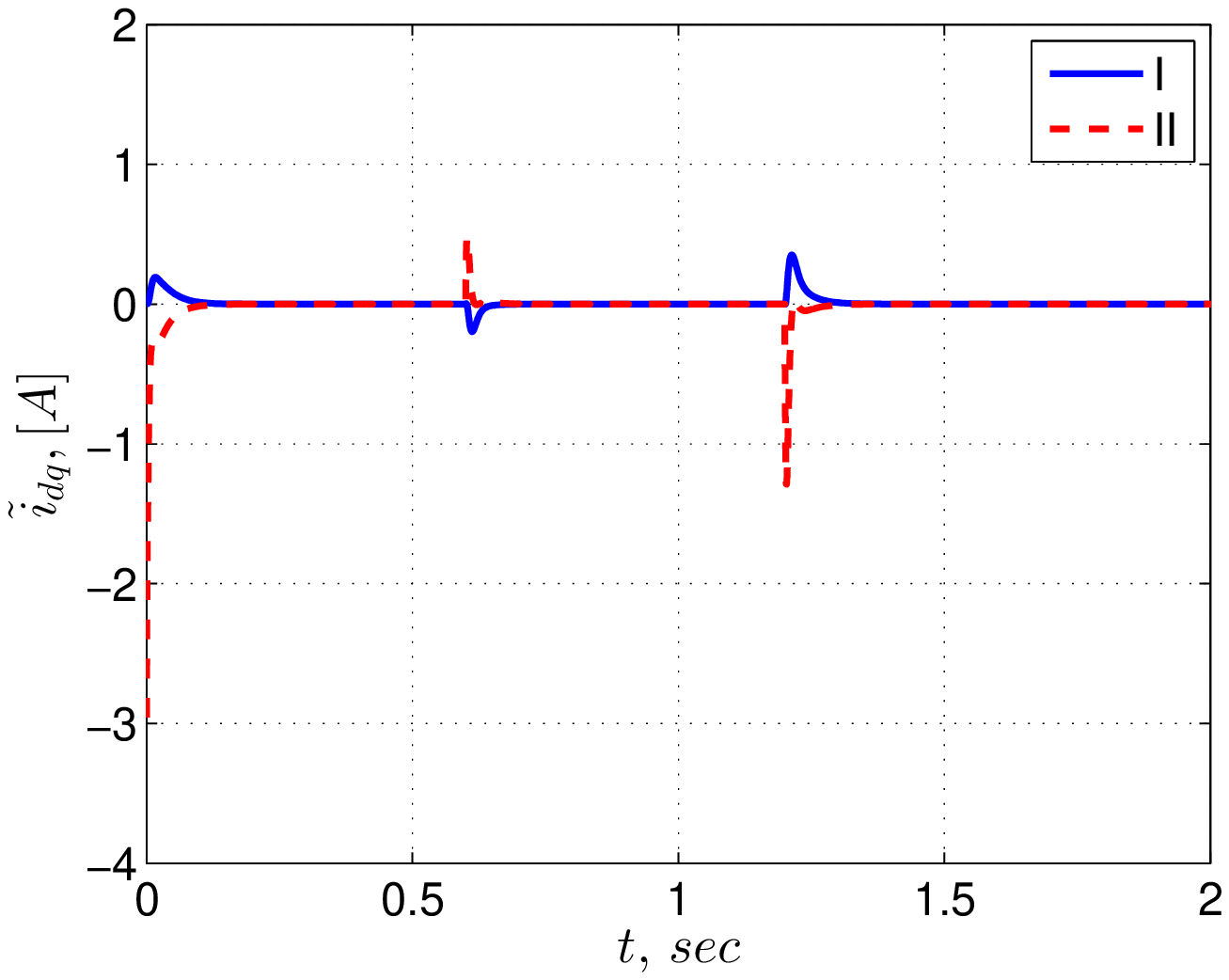}}
\
\subfloat[][Voltages $v_d$ (I) and $v_q$ (II)]{{\label{fig_6c}}\includegraphics[width=0.31\textwidth]{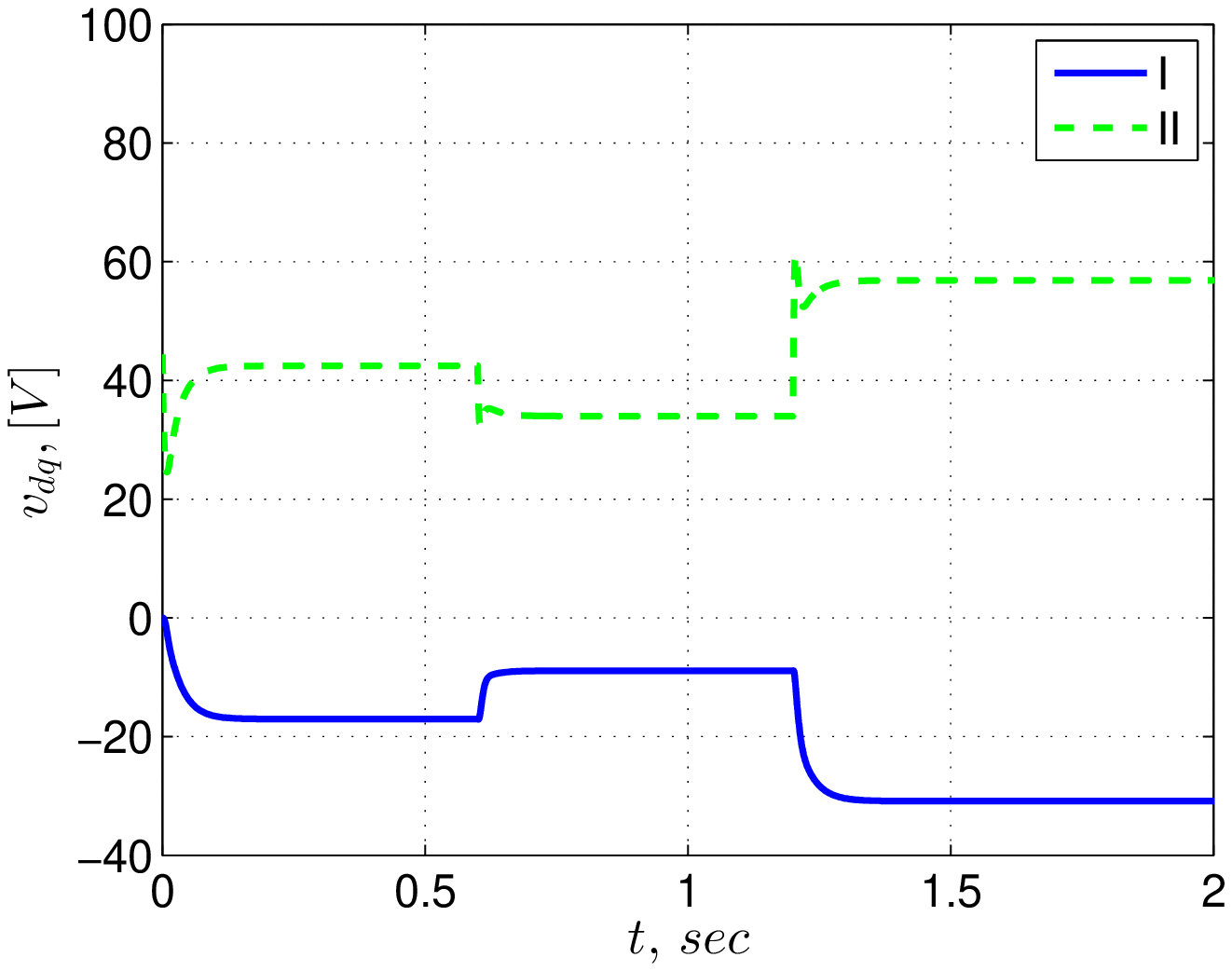}}
\\
\subfloat[][Torque load error]{{\label{fig_6d}}\includegraphics[width=0.31\textwidth]{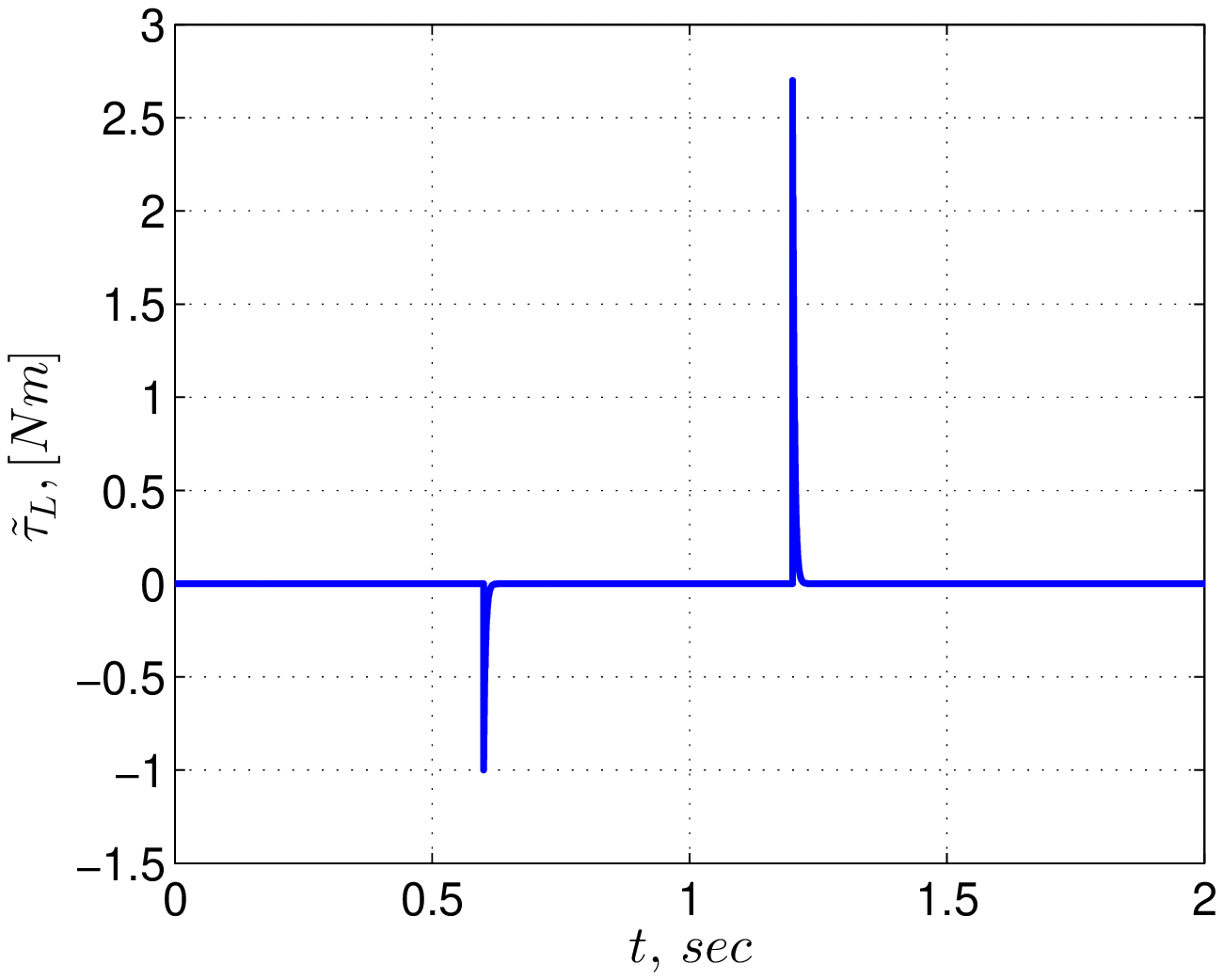}}
\vspace{-2mm}
\caption{Transients in the system with adaptive current PI \eqref{pice}, \eqref{tauobs} and load torque of Fig. \ref{fig_2}: $k_p=15$, $k_i=2000$, $\ell=0.1$}
\label{fig_6}
\end{figure*}

\begin{figure*}[htp]
\centering
\subfloat[][Speed error]{{\label{fig_7a}}\includegraphics[width=0.31\textwidth]{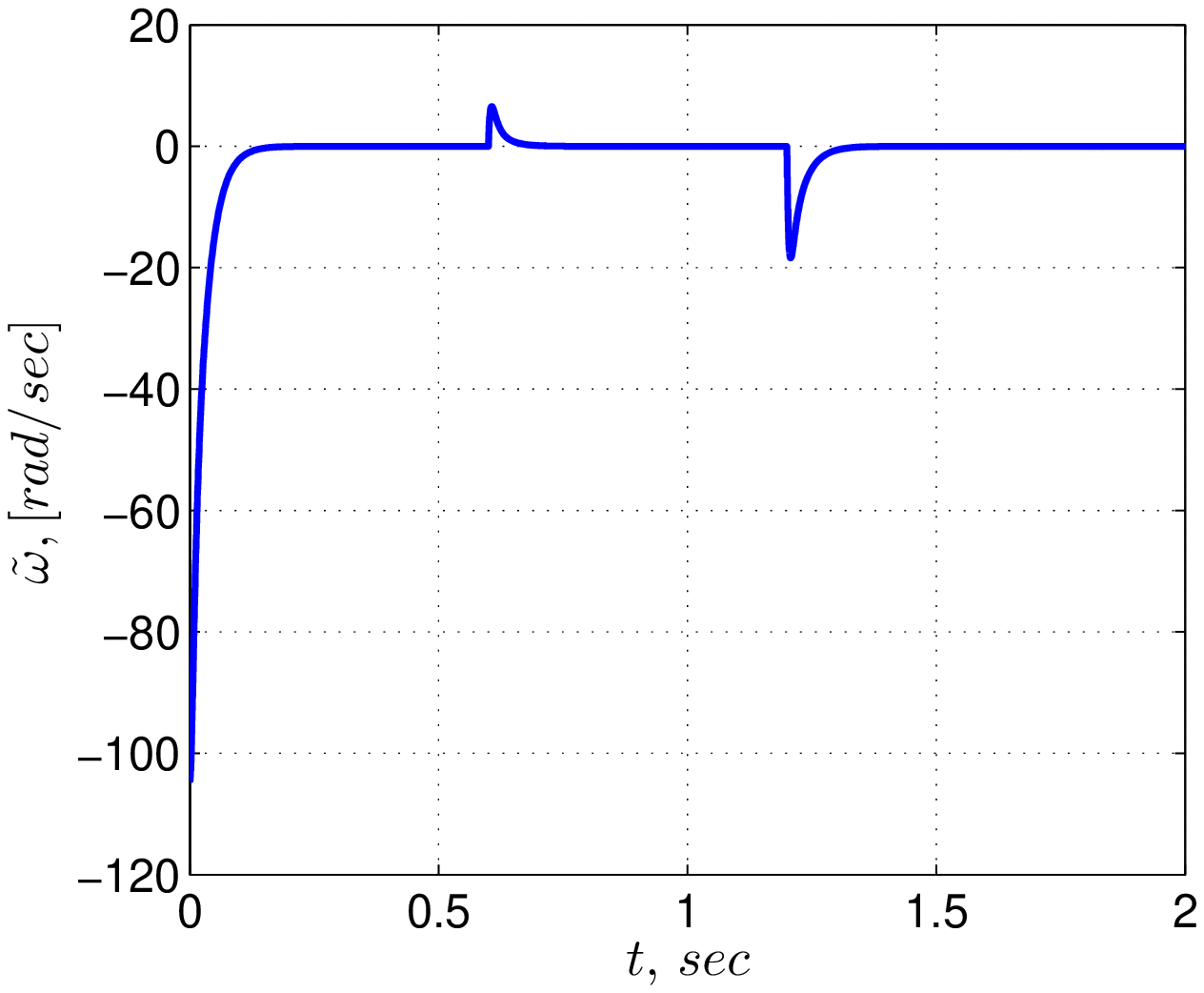}}
\
\subfloat[][Current errors $\tilde{i}_d$ (I) and $\tilde{i}_q$ (II)]{{\label{fig_7b}}\includegraphics[width=0.31\textwidth]{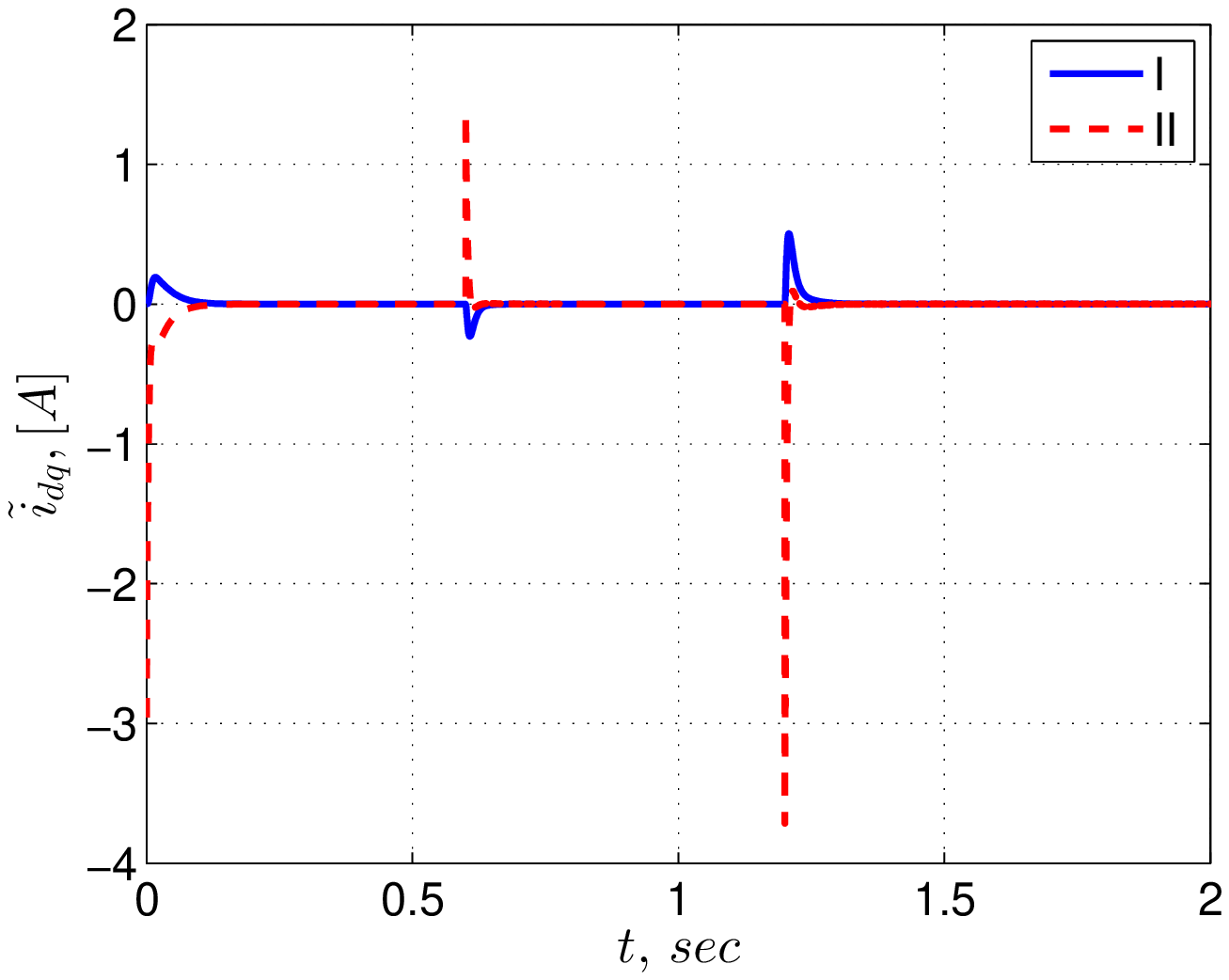}}
\
\subfloat[][Voltages $v_d$ (I) and $v_q$ (II)]{{\label{fig_7c}}\includegraphics[width=0.31\textwidth]{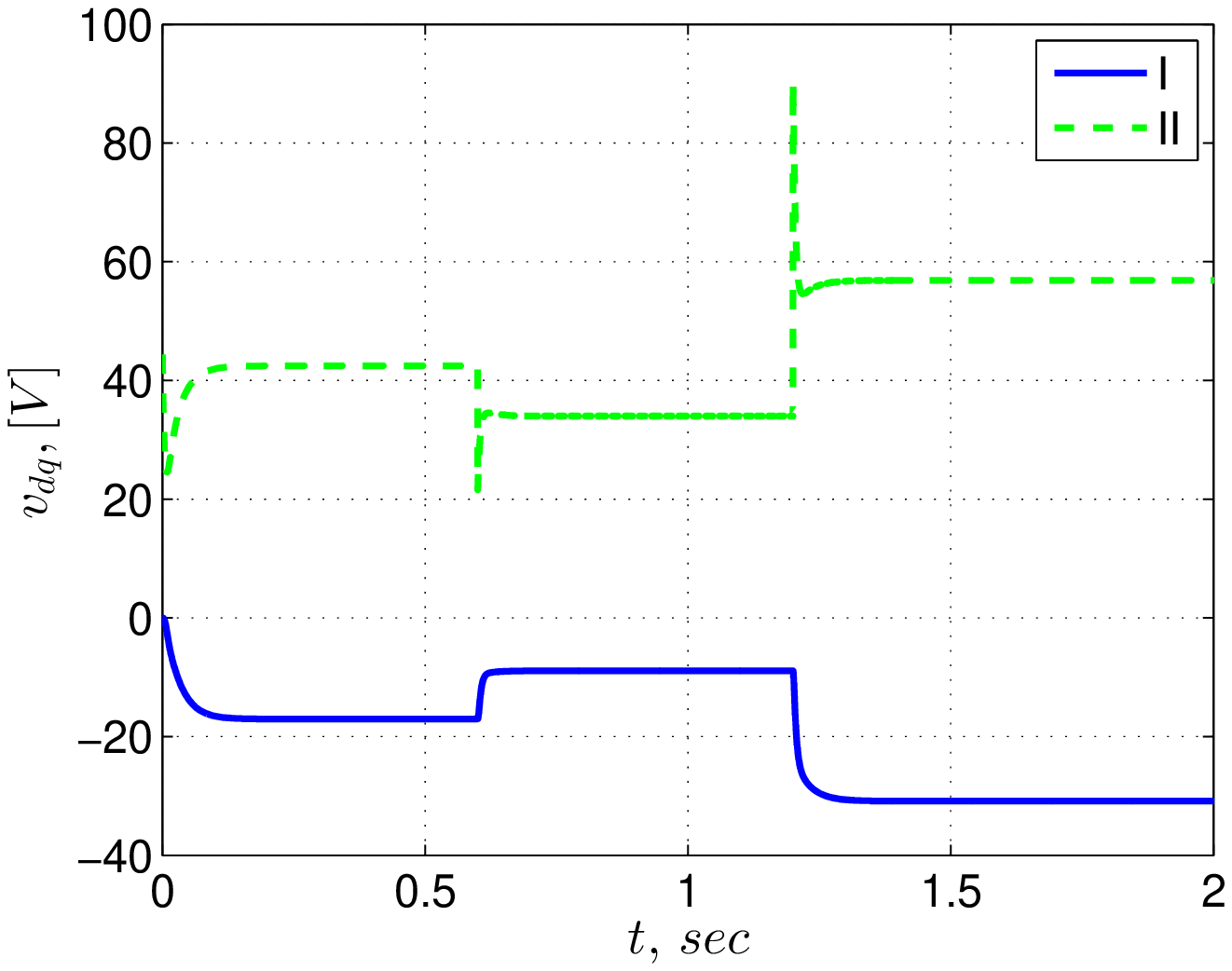}}
\\
\subfloat[][Torque load error]{{\label{fig_7d}}\includegraphics[width=0.31\textwidth]{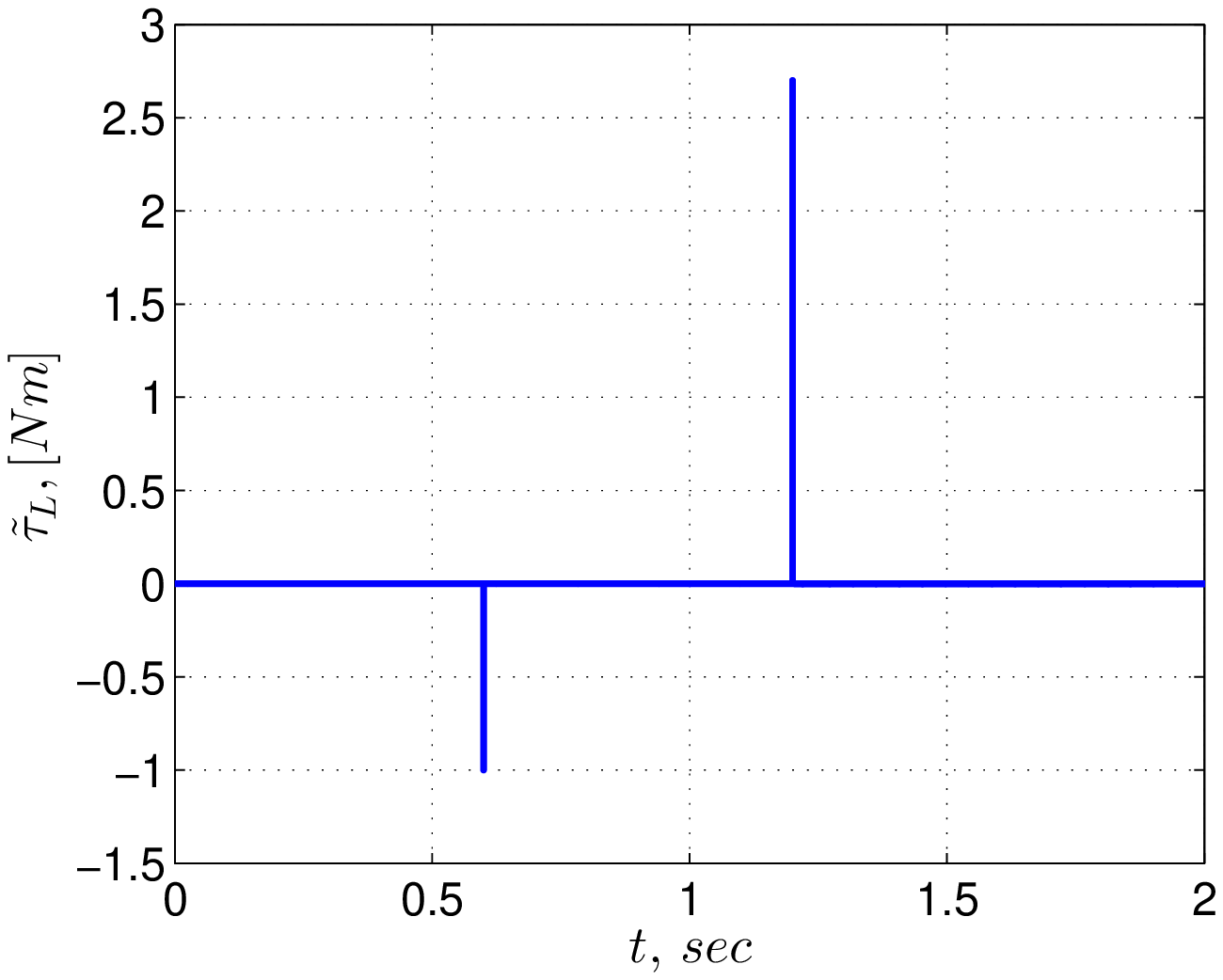}}
\vspace{-2mm}
\caption{Transients in the system with adaptive current PI \eqref{pice}, \eqref{tauobs} and load torque of Fig. \ref{fig_2}: $k_p=15$, $k_i=2000$, $\ell=20$}
\label{fig_7}
\end{figure*}

\begin{figure*}[htp]
\centering
\subfloat[][Speed error]{{\label{fig_8a}}\includegraphics[width=0.31\textwidth]{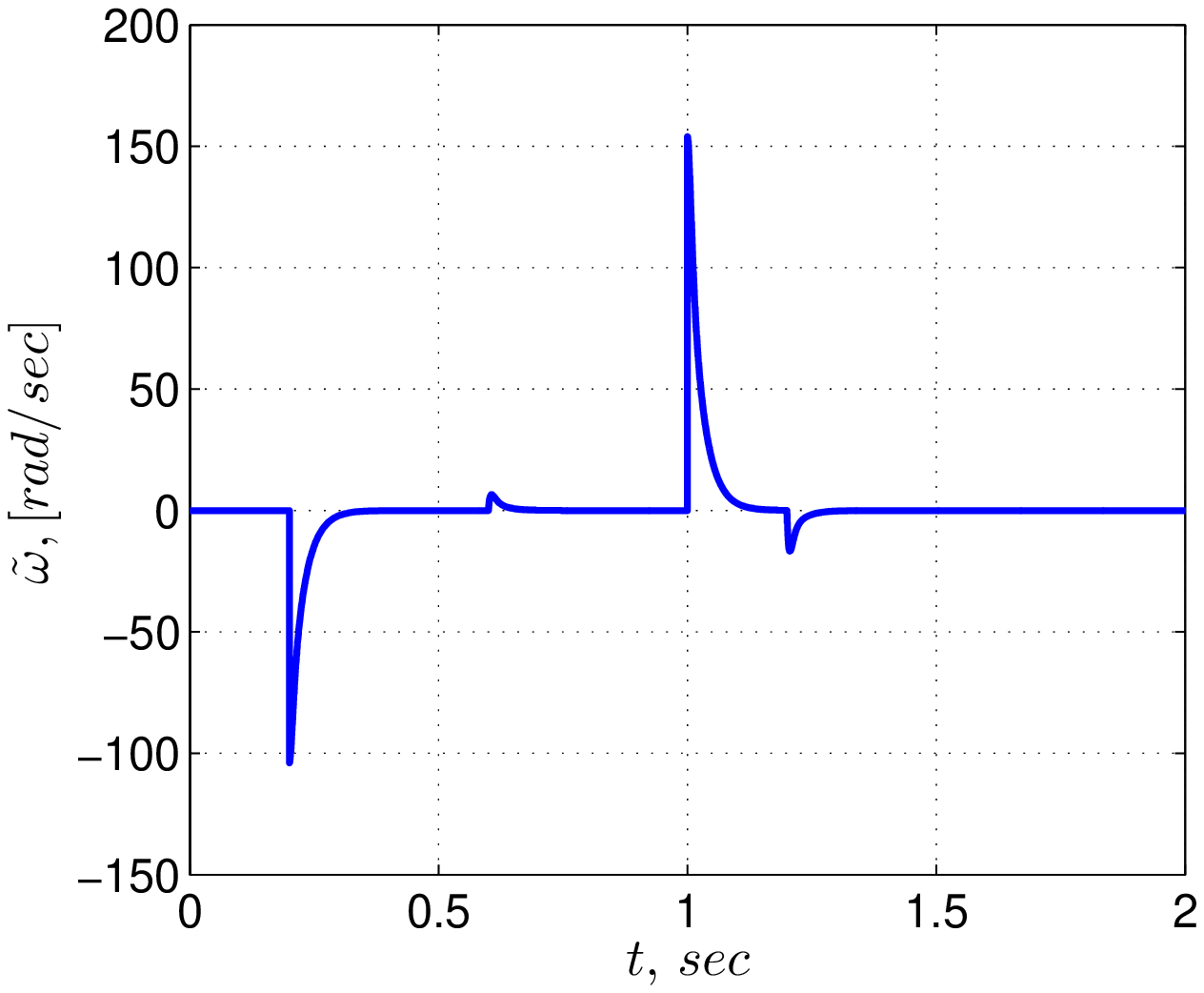}}
\
\subfloat[][Current errors $\tilde{i}_d$ (I) and $\tilde{i}_q$ (II)]{{\label{fig_8b}}\includegraphics[width=0.31\textwidth]{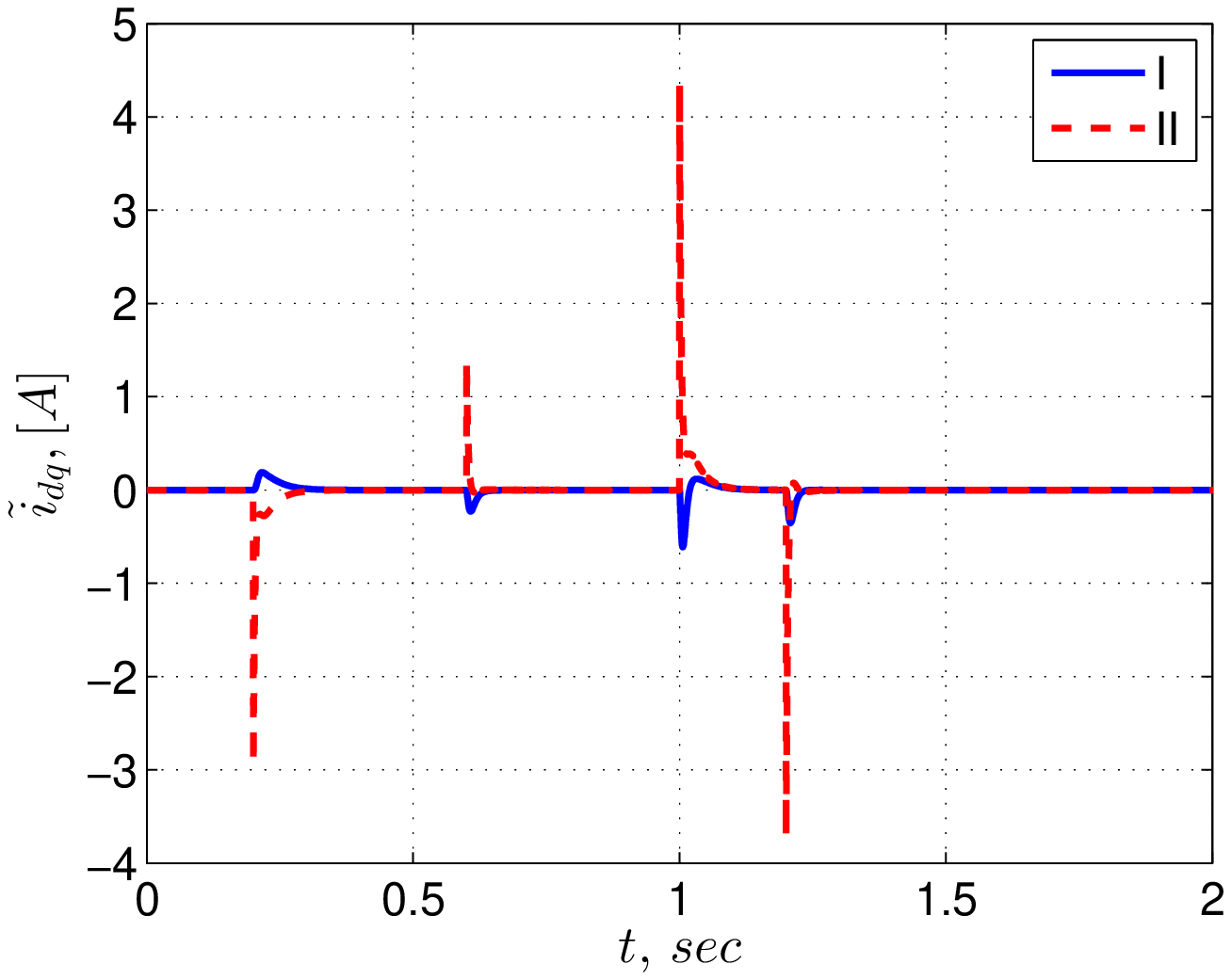}}
\
\subfloat[][Voltages $v_d$ (I) and $v_q$ (II)]{{\label{fig_8c}}\includegraphics[width=0.31\textwidth]{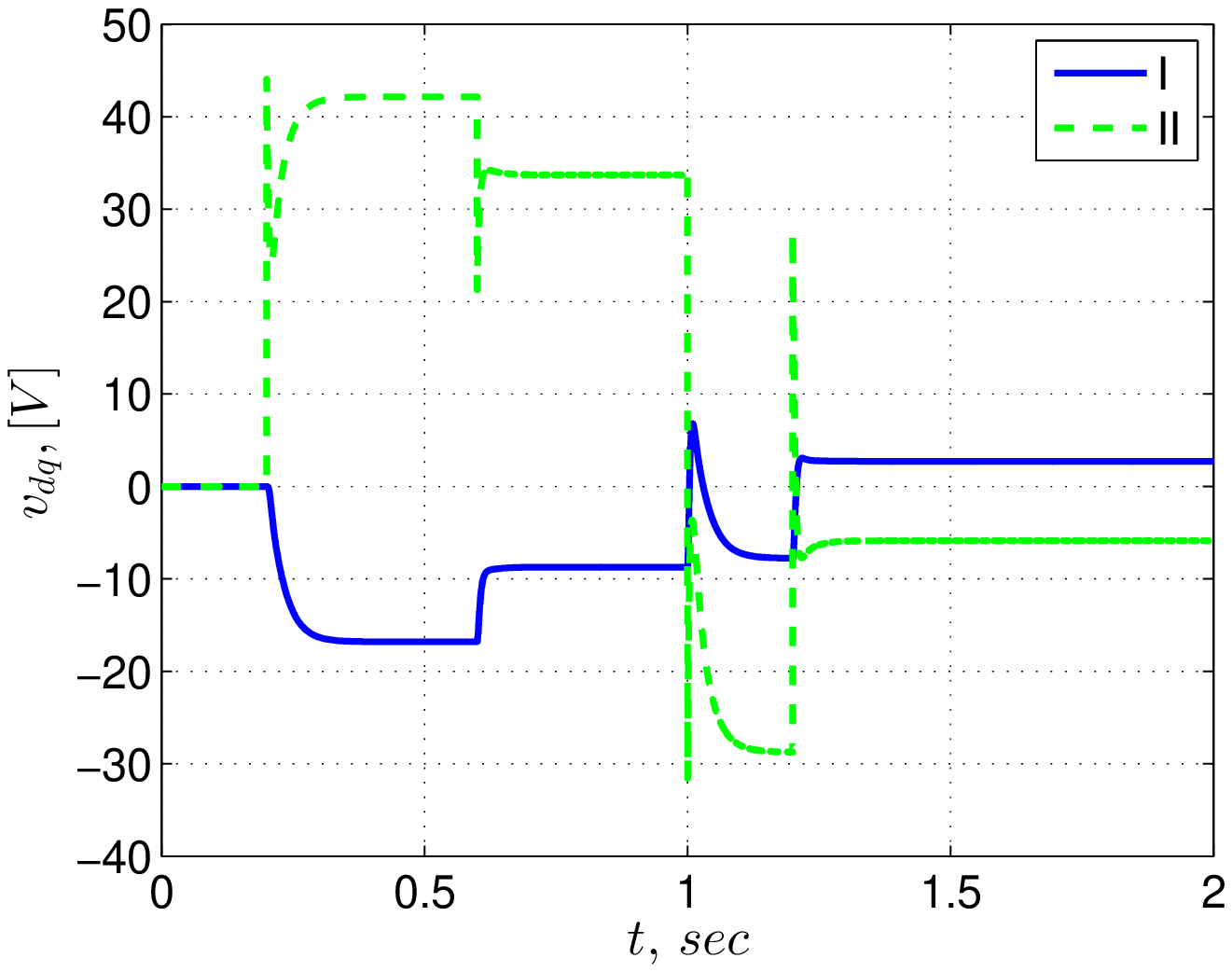}}
\\
\subfloat[][Torque load error]{{\label{fig_8d}}\includegraphics[width=0.31\textwidth]{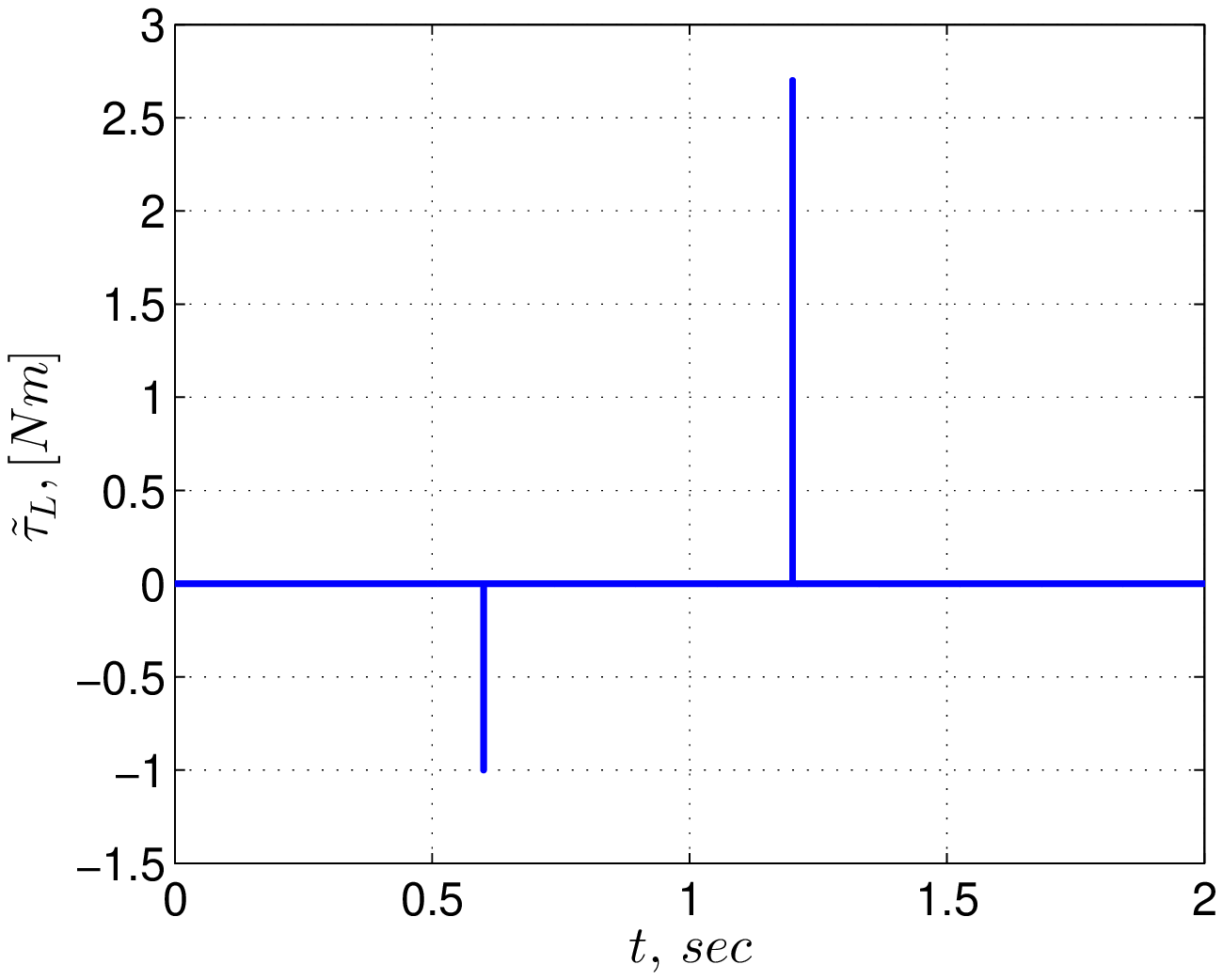}}
\vspace{-2mm}
\caption{Transients in the system with adaptive current PI \eqref{pice}, \eqref{tauobs},  load torque of Fig. \ref{fig_2} and speed reference of Fig. \ref{fig_1}: $k_p=15$, $k_i=2000$, $\ell=20$}
\label{fig_8}
\end{figure*}

The transients for the adaptive PI with load torque estimator \eqref{tauobs} are shown in Figs. \ref{fig_6} and \ref{fig_7}. Here we use the time-varying torque profile of Fig. \ref{fig_2}. In Fig. \ref{fig_7} the gain $\ell$ is taken higher than in Fig. \ref{fig_6}, that as expected from \eqref{dynetau}, leads to a faster convergence of $\tilde{\tau}_L$ to zero and smaller speed errors. The next test, shown in Fig. \ref{fig_8}, illustrates the system behaviour for the varying speed reference given in Fig. \ref{fig_1} with the same load input and gains as in the previous scenario. As seen from the figure the adaptive PI controller provides good performance and all the errors $\tilde{\tau}_L$, $\tilde{\omega}$ and $\tilde{i}_{dq}$ converge to zero fast.

\begin{figure*}[htp]
\centering
\subfloat[][Speed error]{{\label{fig_9a}}\includegraphics[width=0.31\textwidth]{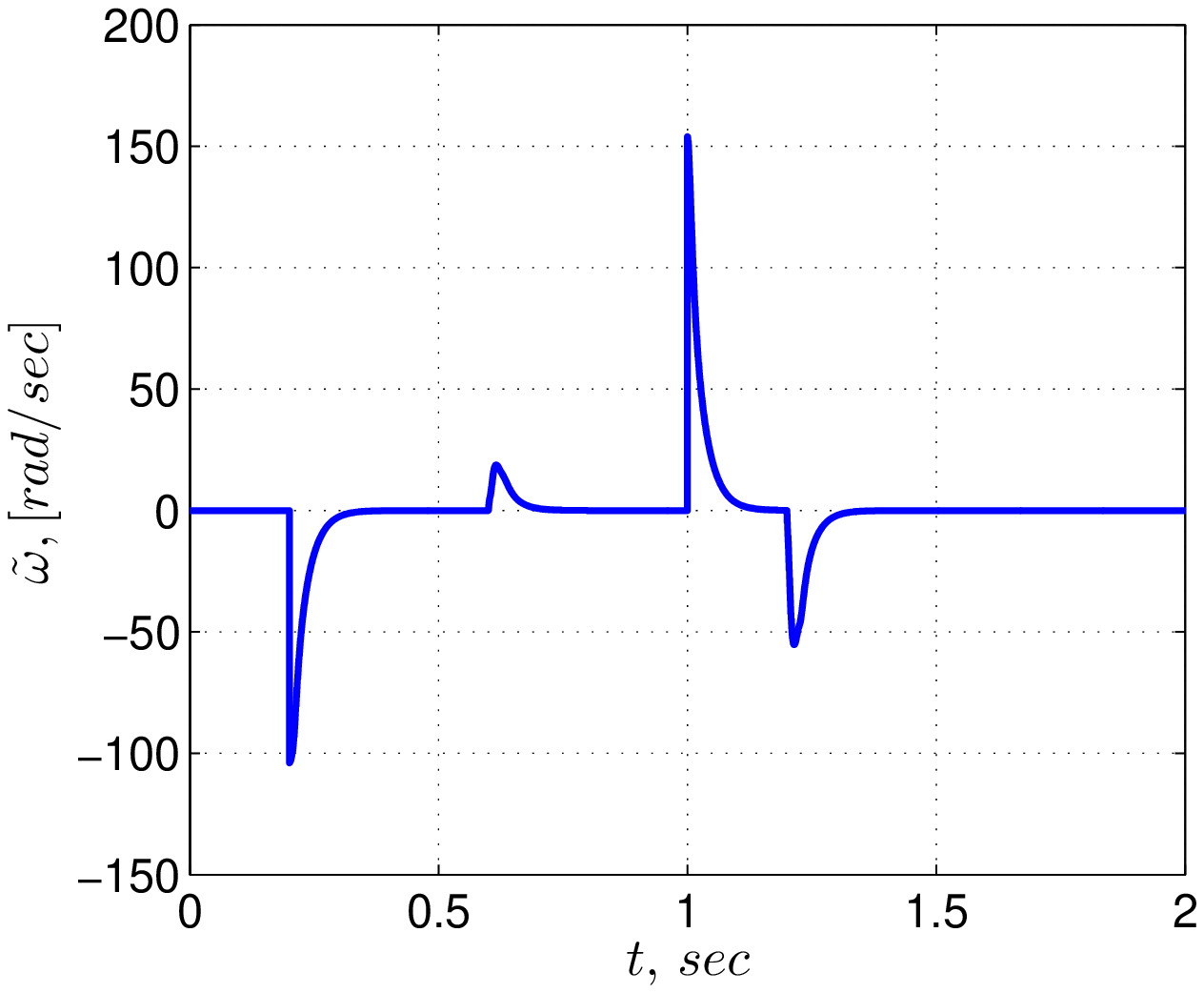}}
\
\subfloat[][Current errors $\tilde{i}_d$ (I) and $\tilde{i}_q$ (II)]{{\label{fig_9b}}\includegraphics[width=0.31\textwidth]{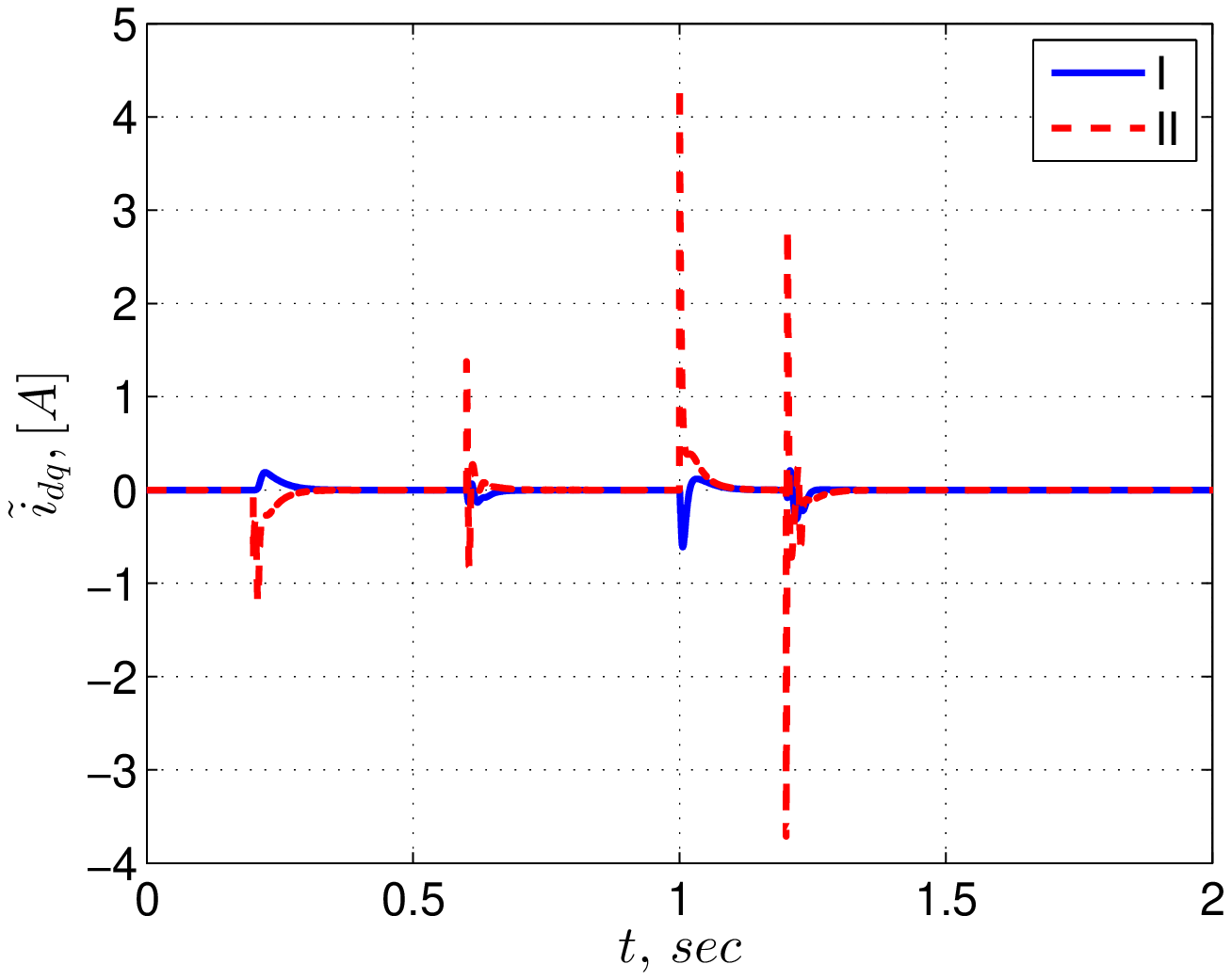}}
\
\subfloat[][Voltages $v_d$ (I) and $v_q$ (II)]{{\label{fig_9c}}\includegraphics[width=0.31\textwidth]{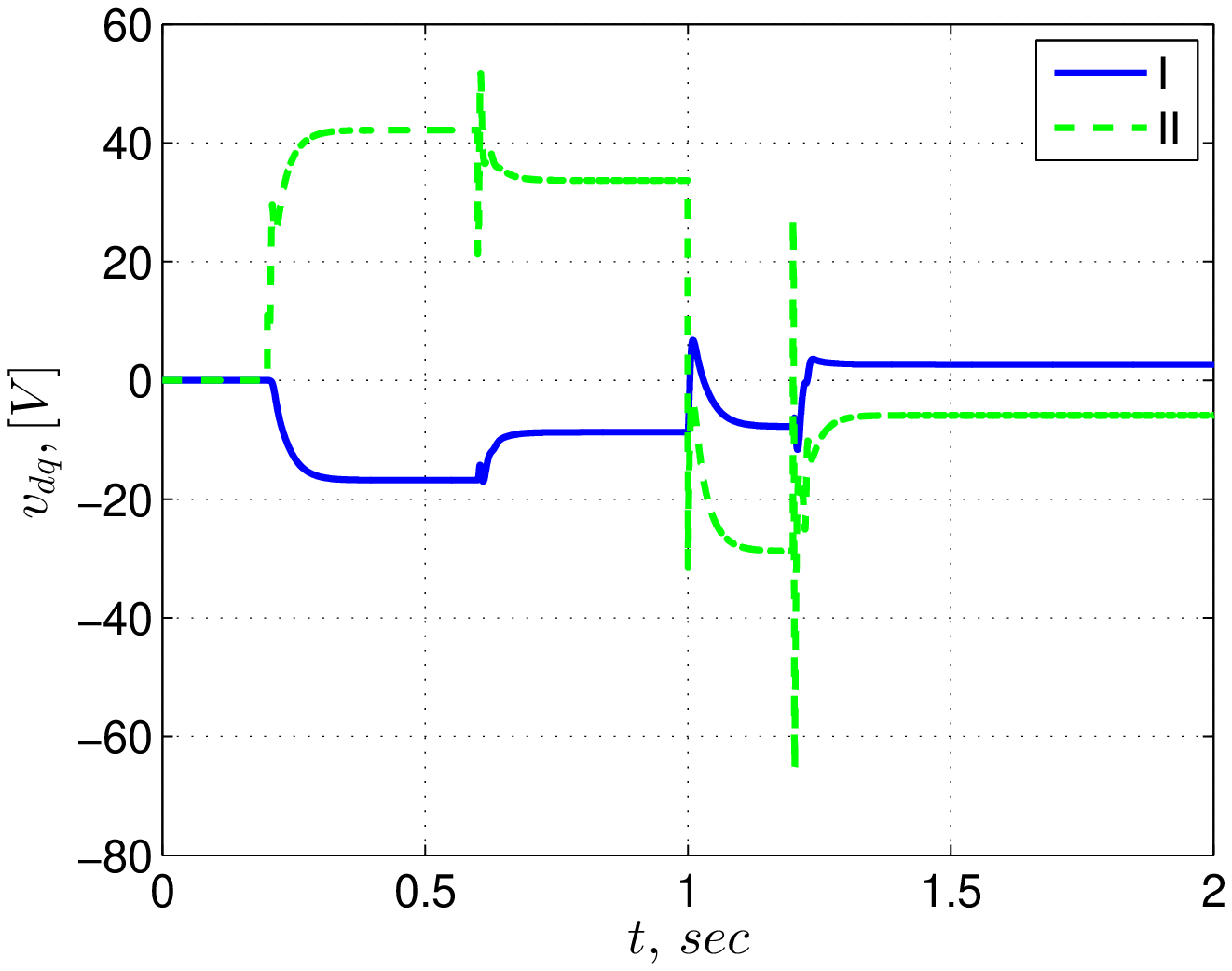}}
\\
\subfloat[][Torque load error]{{\label{fig_9d}}\includegraphics[width=0.31\textwidth]{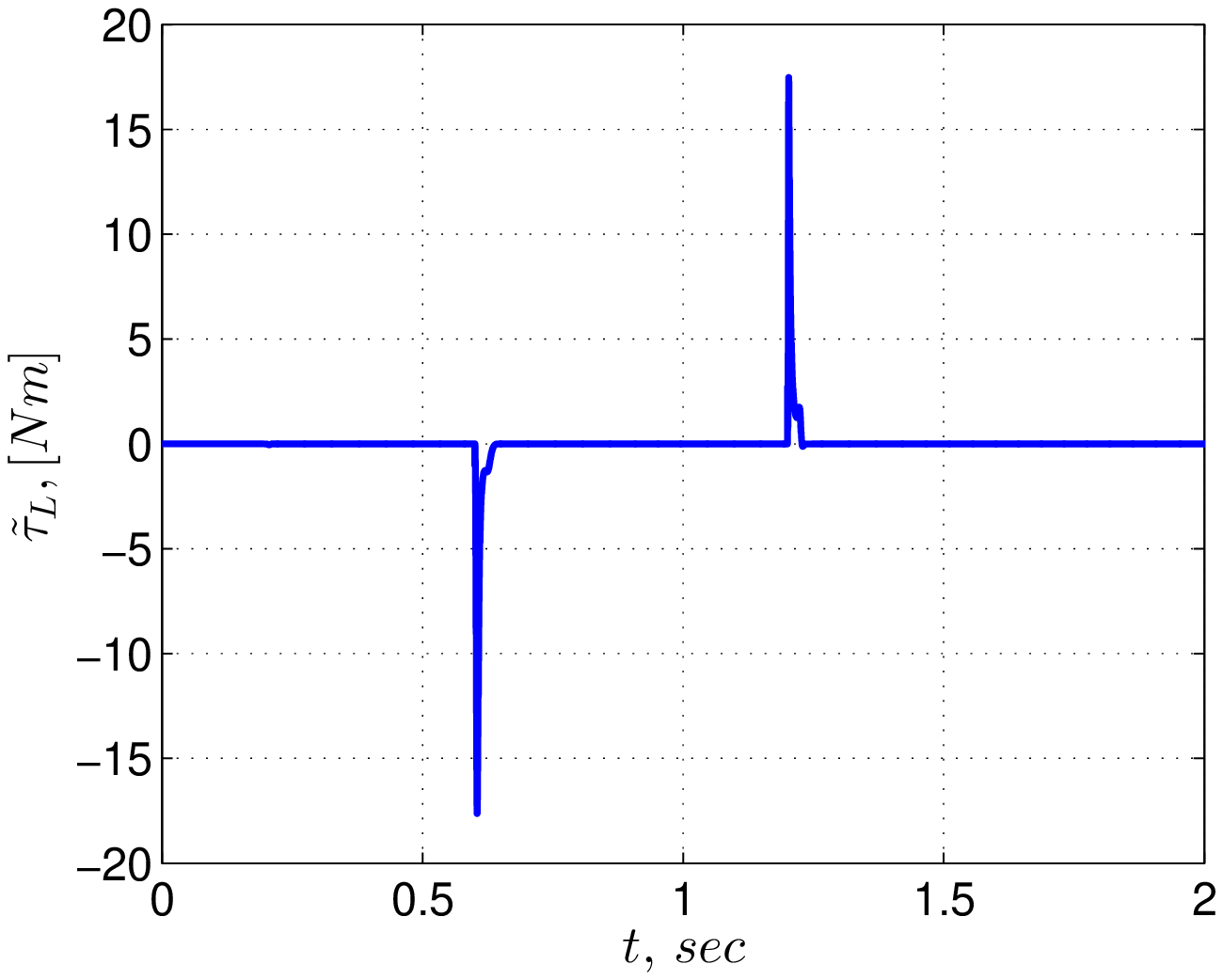}}
\
\subfloat[][$R_m$ estimation error]{{\label{fig_9e}}\includegraphics[width=0.31\textwidth]{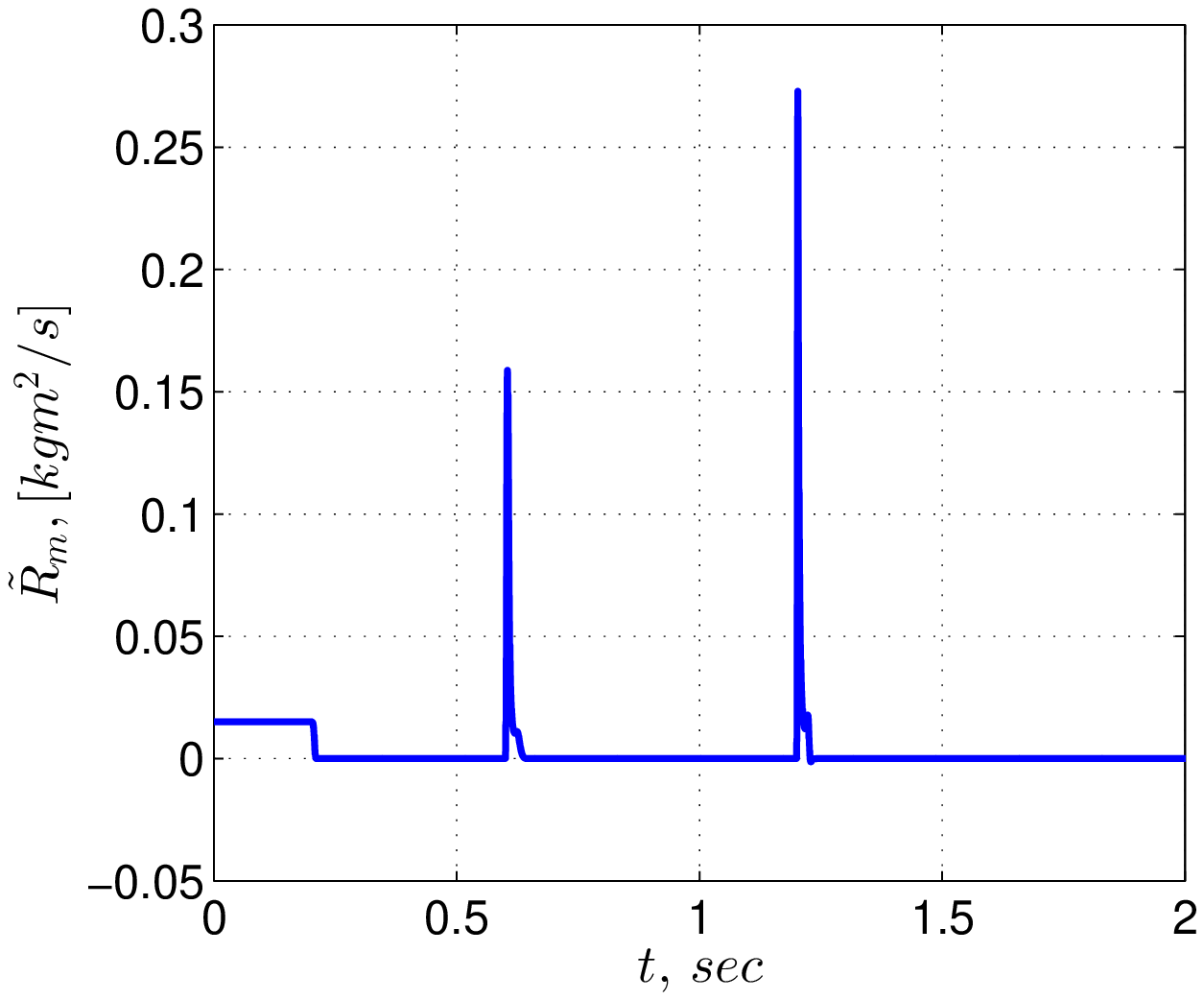}}
\vspace{-2mm}
\caption{Transients in the system with adaptive current PI \eqref{pice} with estimators of load torque \eqref{tauobs} and resistance \eqref{estrm}, load torque of Fig. \ref{fig_2} and speed reference of Fig. \ref{fig_1}: $k_p=15$, $k_i=2000$, $\ell=20$, $\alpha=\beta=300$, $\gamma=200$, $\hat R_m(0)=0.005$ Nm}
\label{fig_9}
\end{figure*}

\begin{figure*}[htp]
\centering
\subfloat[][Speed error]{{\label{fig_10a}}\includegraphics[width=0.31\textwidth]{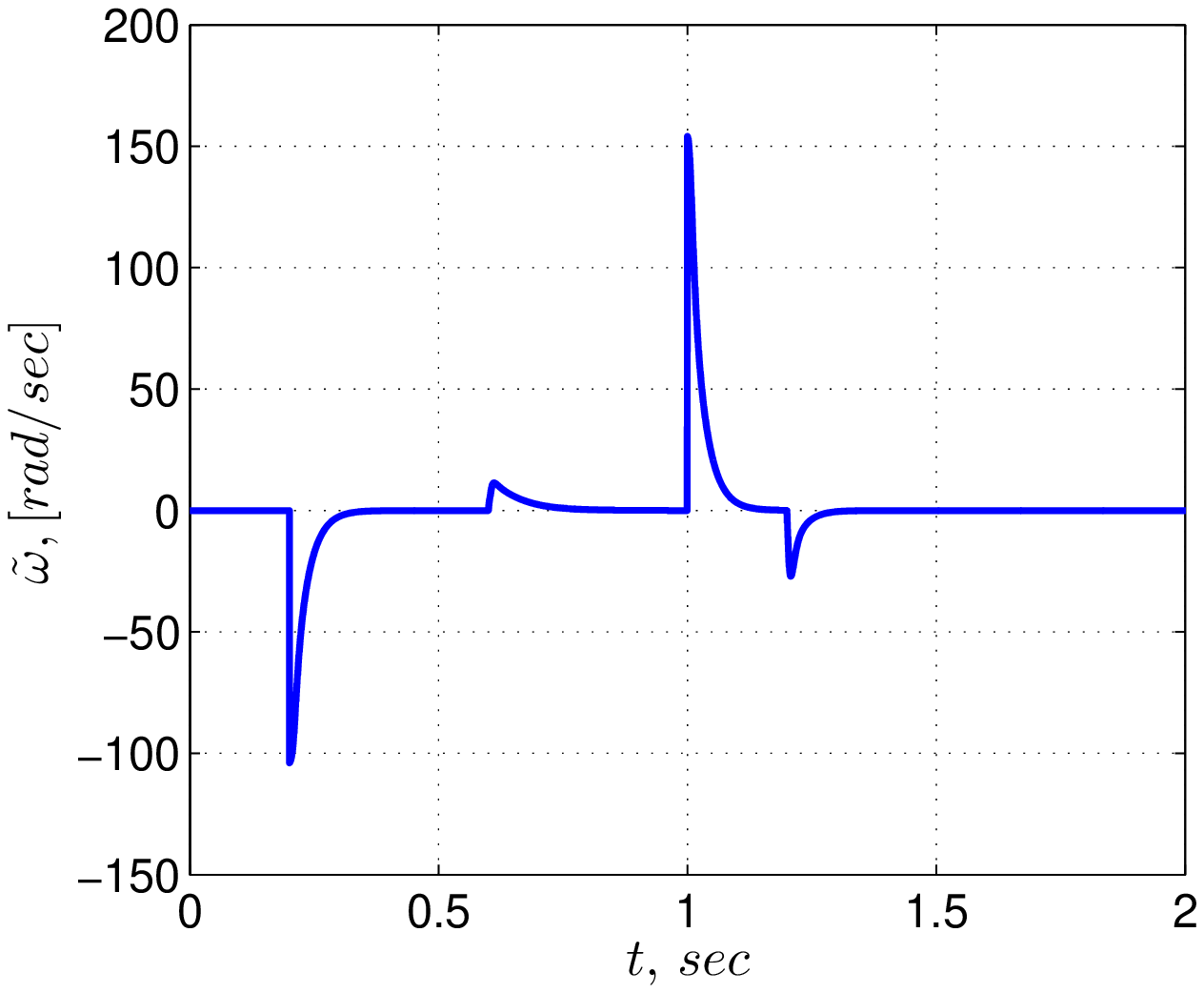}}
\
\subfloat[][Current errors $\tilde{i}_d$ (I) and $\tilde{i}_q$ (II)]{{\label{fig_10b}}\includegraphics[width=0.31\textwidth]{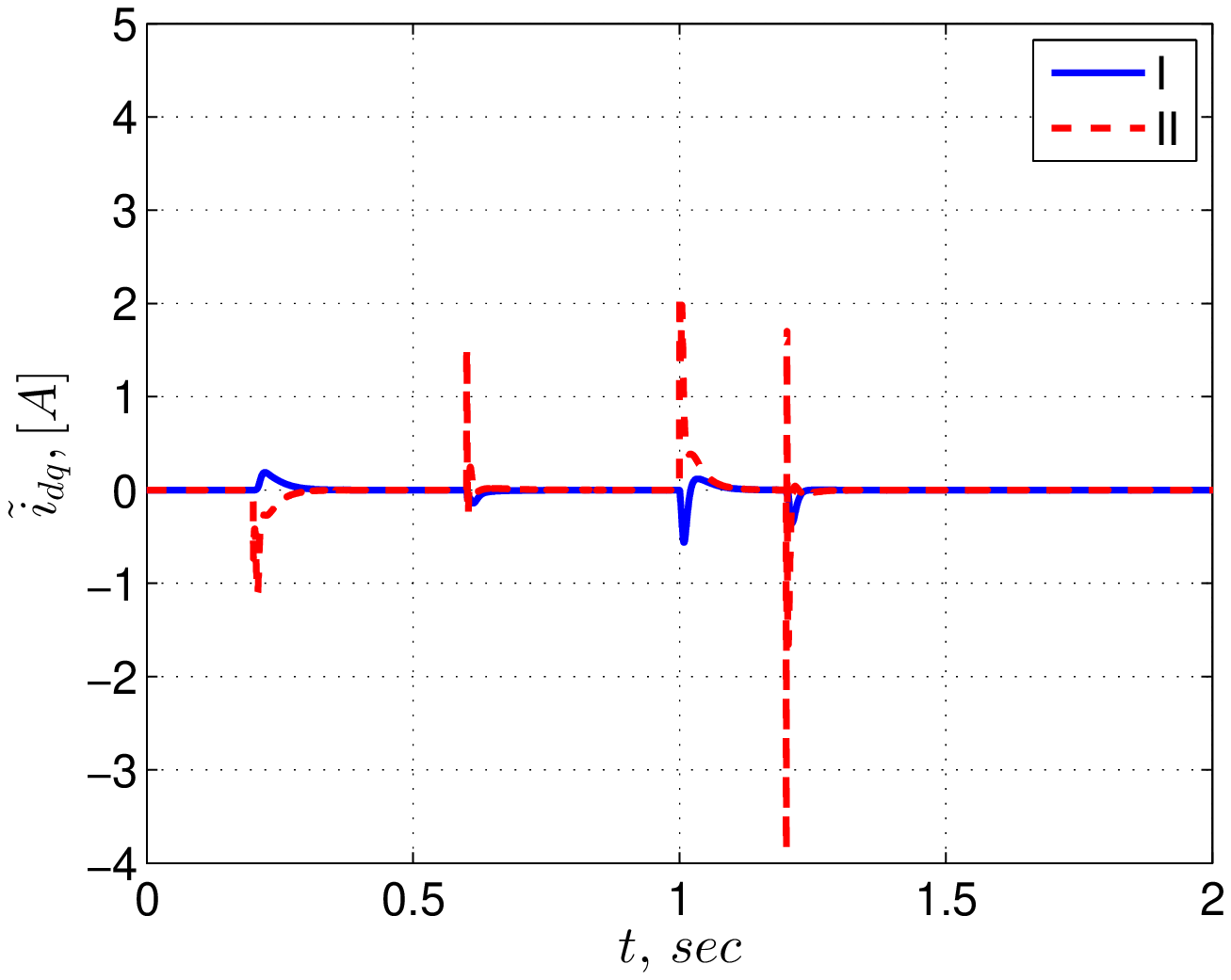}}
\
\subfloat[][Voltages $v_d$ (I) and $v_q$ (II)]{{\label{fig_10c}}\includegraphics[width=0.31\textwidth]{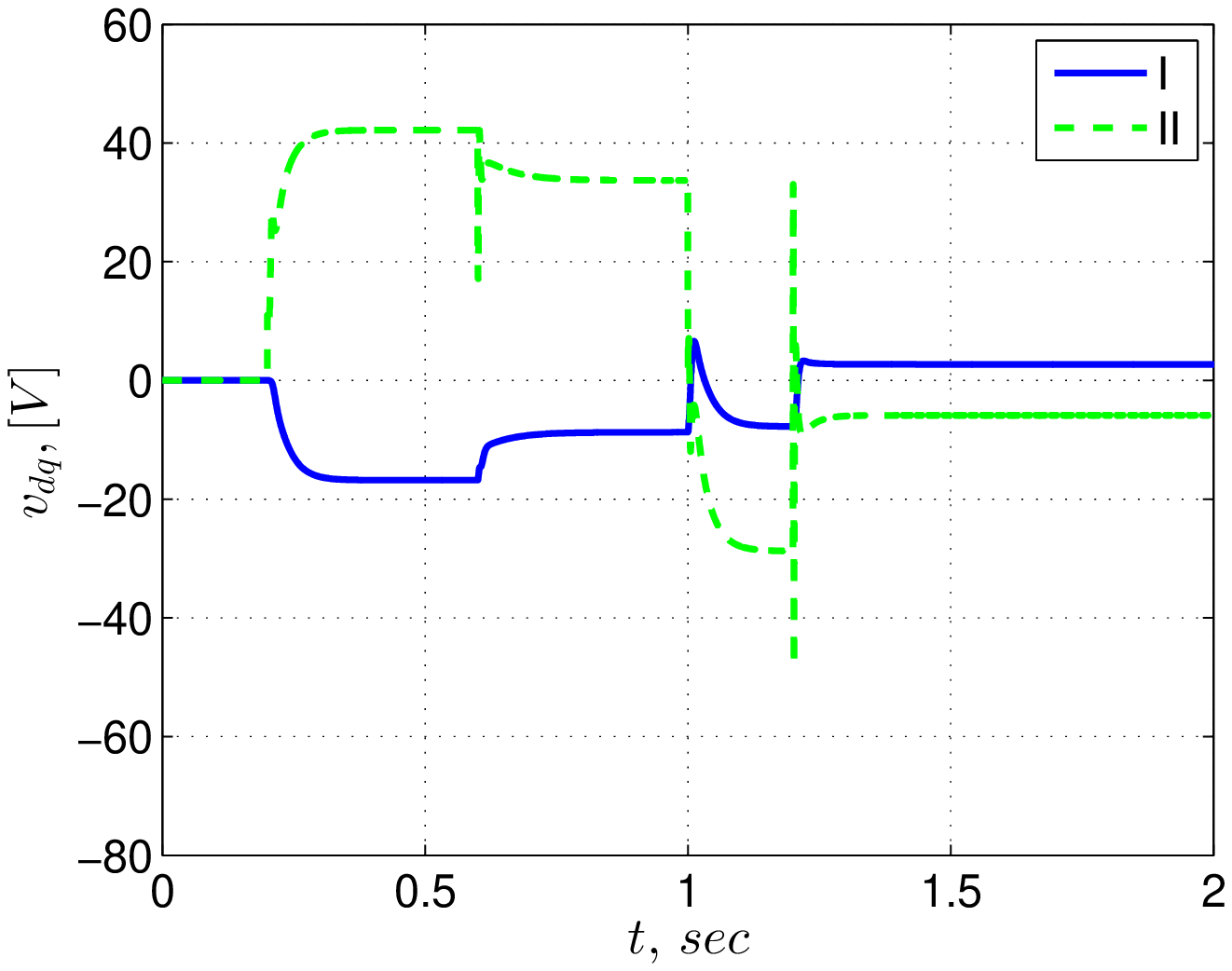}}
\\
\subfloat[][Torque load error]{{\label{fig_10d}}\includegraphics[width=0.31\textwidth]{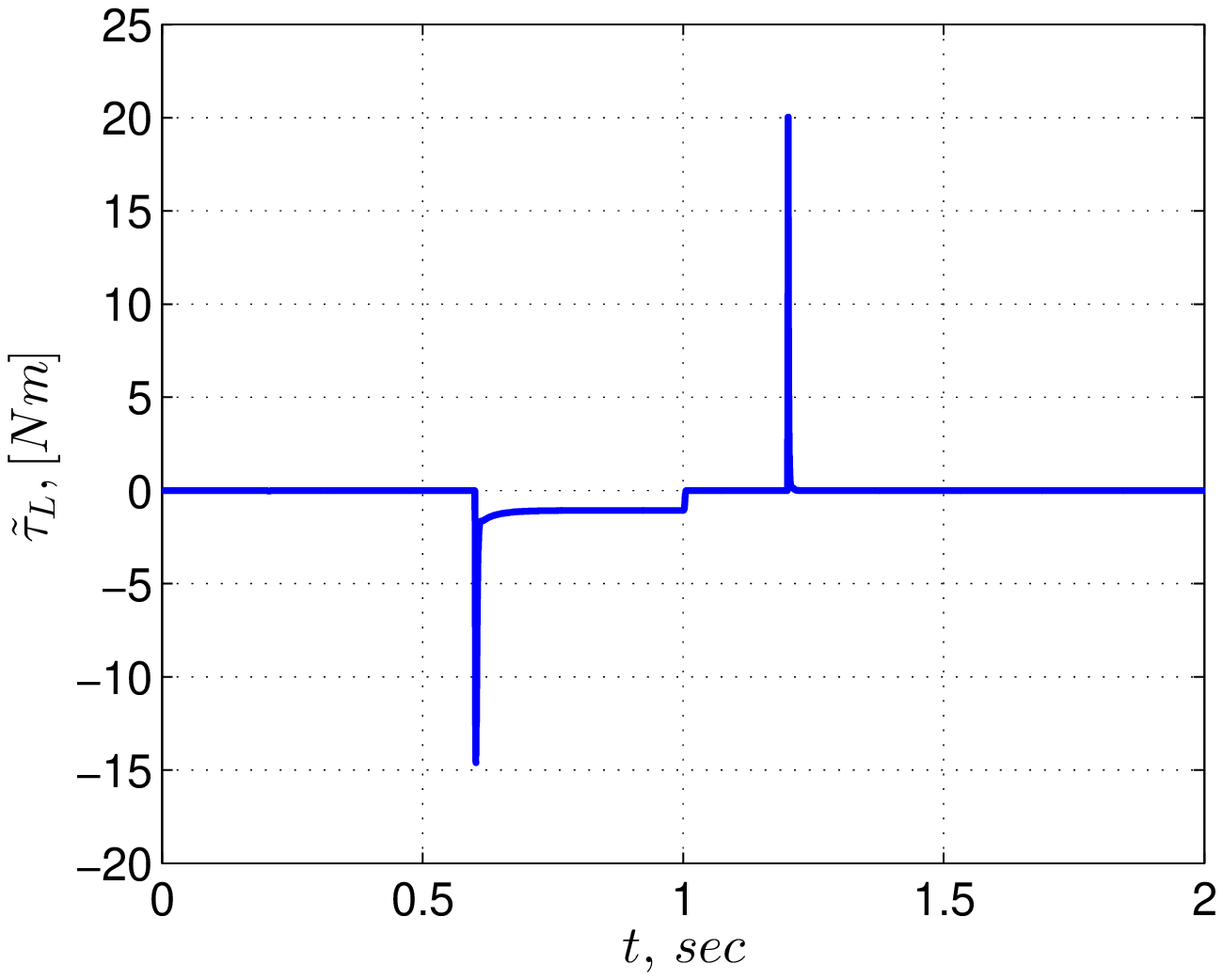}}
\
\subfloat[][$R_m$ estimation error]{{\label{fig_10e}}\includegraphics[width=0.31\textwidth]{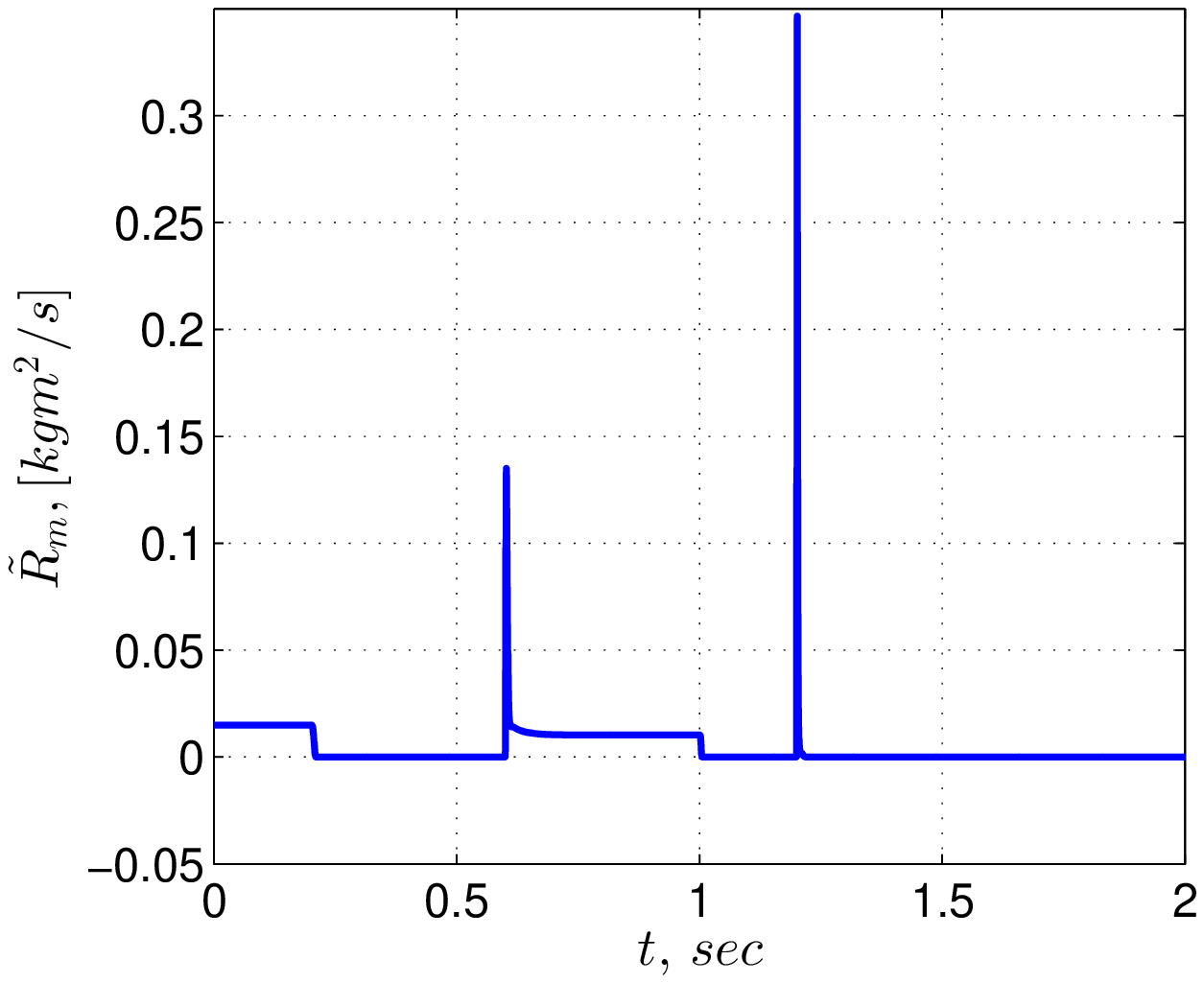}}
\vspace{-2mm}
\caption{Transients in the system with adaptive current PI  \eqref{pice} with estimators of load torque \eqref{tauobs} and resistance \eqref{estrm},  load torque of Fig. \ref{fig_2} and speed reference of Fig. \ref{fig_1}: $k_p=15$, $k_i=2000$, $\ell=10$, $\alpha=\beta=1300$, $\gamma=500$, $\hat R_m(0)=0.005$ Nm}
\label{fig_10}
\end{figure*}

Simulations for the adaptive PI \eqref{pice}, \eqref{tauobs} equipped with the estimate of viscous friction coefficient $\hat{R}_m$ generated by \eqref{estrm} are shown in Figs. \ref{fig_9} and \ref{fig_10} for two different gain settings and the time-varying profiles of Figs \ref{fig_1} and \ref{fig_2}. In both cases $\hat{R}_m(0)=0.005$, which is $25\%$ of the actual value. While the choice of parameters of Fig. \ref{fig_9} is suitable for the simultaneous estimation of ${\tau}_L$ and ${R}_m$, reducing the constant $\ell$ and changing the bandwidth of the filter ${\beta p \over (p+\alpha)^2}$ has a deleterious effect. Indeed, as shown in Fig. \ref{fig_10} there is a static error in both estimators in the interval $t \in (0.6,1)$ sec., and it is not until the appearance of the speed reference change at $t=1$ sec. that the estimators recover their alertness. This observation underscores, on one hand, the need of excitation indicated in Lemma \ref{lem3} and, on the other hand, the importance of selecting suitable tuning gains for the estimators.  

\begin{figure*}[htp]
\centering
\subfloat[][Speed error]{{\label{fig_11a}}\includegraphics[width=0.31\textwidth]{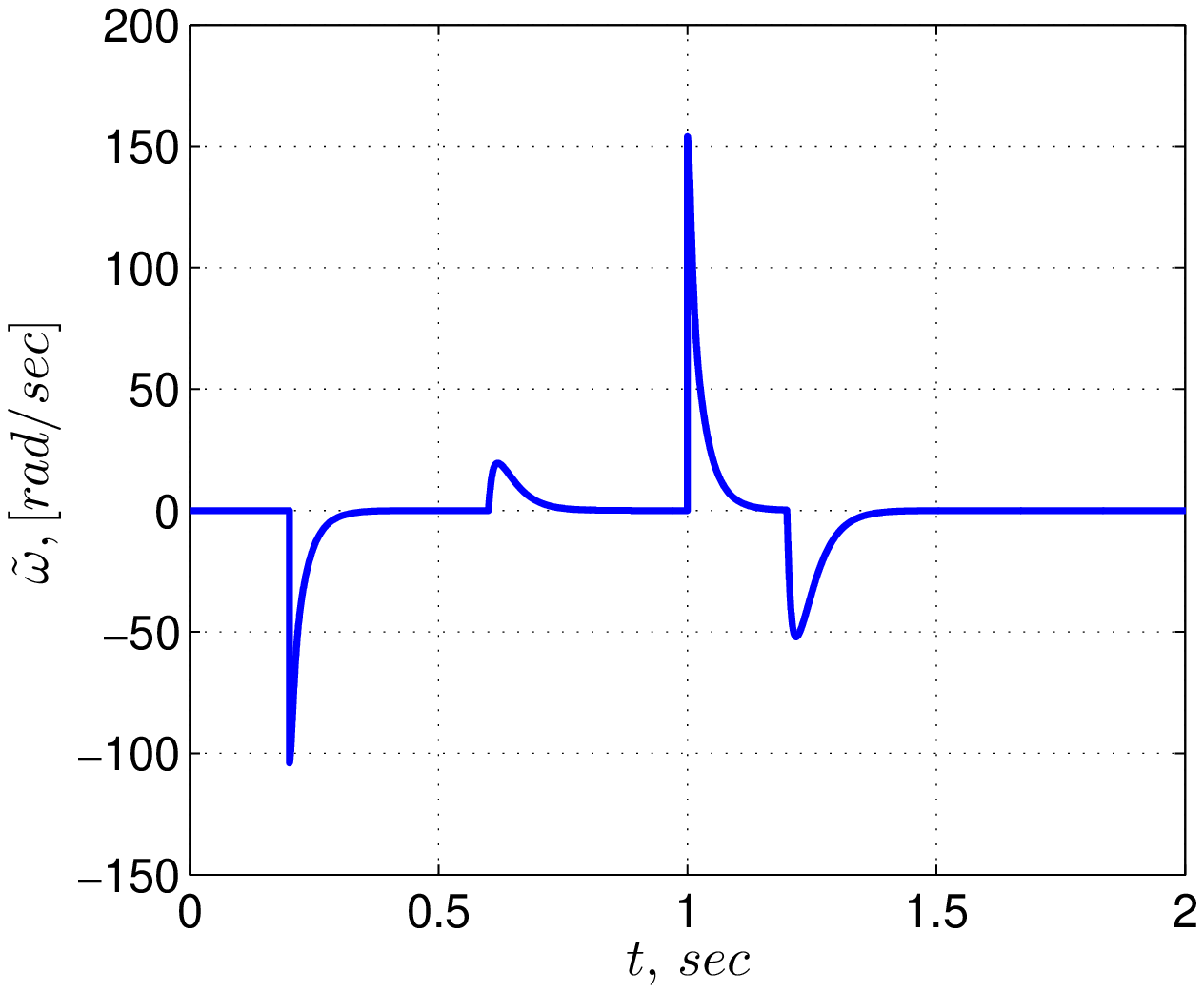}}
\
\subfloat[][Current errors $\tilde{i}_d$ (I) and $\tilde{i}_q$ (II)]{{\label{fig_11b}}\includegraphics[width=0.31\textwidth]{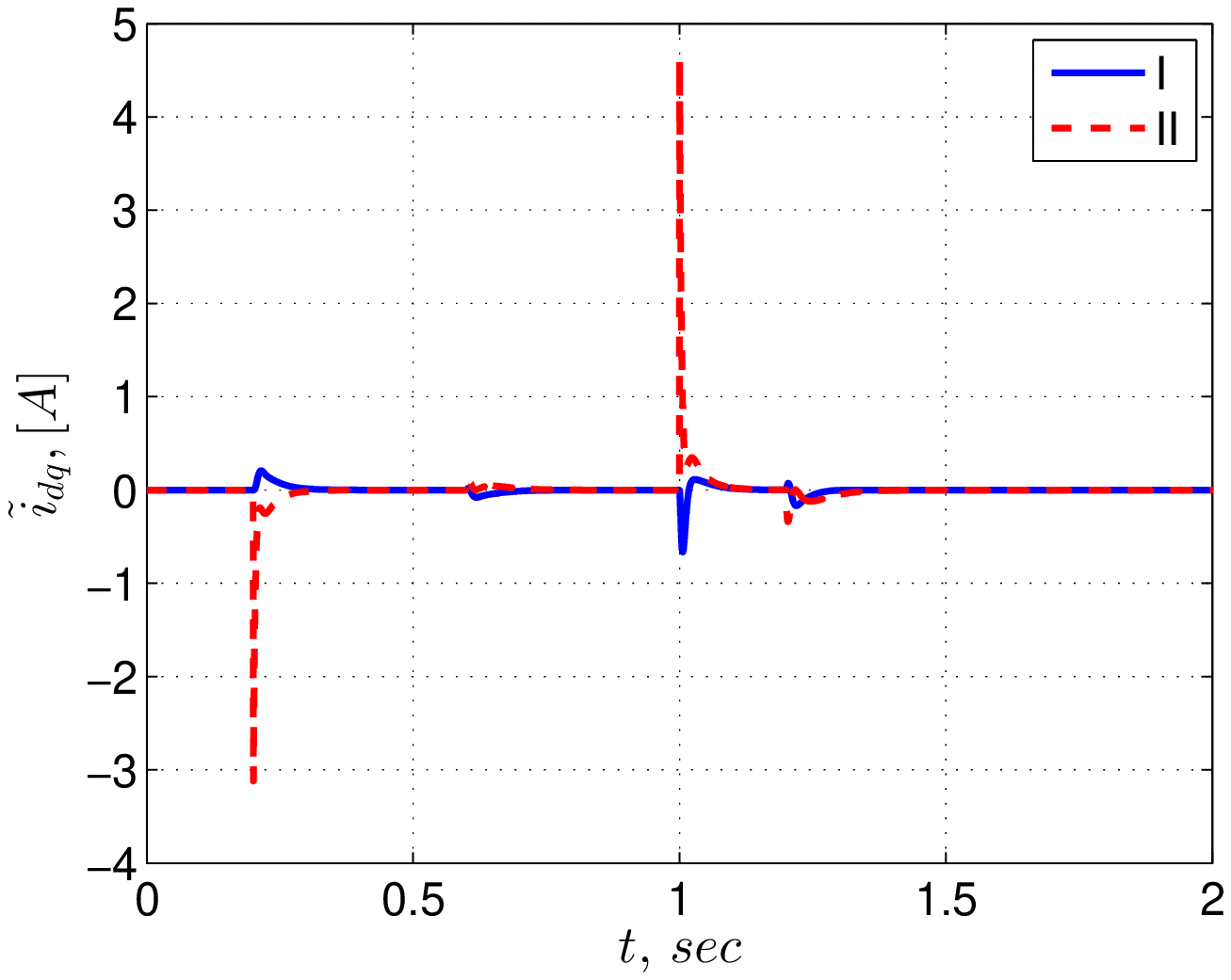}}
\
\subfloat[][Voltages $v_d$ (I) and $v_q$ (II)]{{\label{fig_11c}}\includegraphics[width=0.31\textwidth]{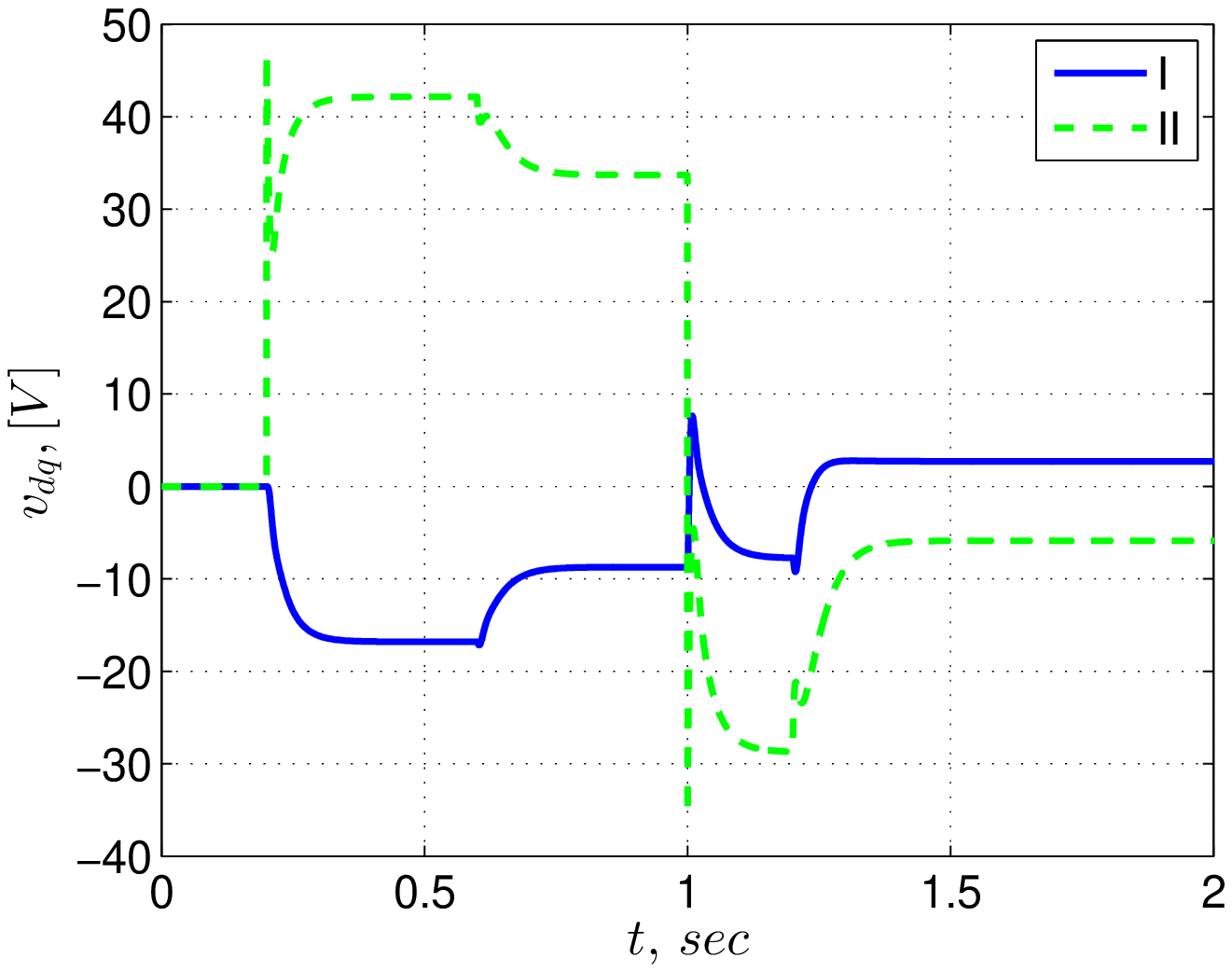}}
\vspace{-2mm}
\caption{Transients in the system with inner-loop current PI \eqref{pi} and outer-loop speed PI \eqref{nespi},  load torque of Fig. \ref{fig_2} and speed reference of Fig. \ref{fig_1}: $k_p=15$, $k_i=2000$, $a_p=0.03$ and $a_i=1.1$.}
\label{fig_11}
\end{figure*}

Fig. \ref{fig_11} illustrates the transients of the system with the standard PI \eqref{pi} and outer-loop PI around speed errors \eqref{nespi} that, as discussed throughout, is often used in practice. In order to compare the efficiency with the adaptive PI we use the same gains $k_p=15$ and $k_i=2000$ for the current regulation as in the previous test (Fig. \ref{fig_9}). The torque load and speed references are also the same. The gains $a_P=a_p I$ and $a_I=a_i I$ of the outer-loop speed controller are tuned to attain similar current and voltage levels. Comparing Figs. \ref{fig_9a} and \ref{fig_11a} one can see that proposed controller with guaranteed GAS ensures a faster speed regulation with lower overshoot when the load changes. 

\begin{figure*}[htp]
\centering
\subfloat[][Speed error]{{\label{fig_12a}}\includegraphics[width=0.31\textwidth]{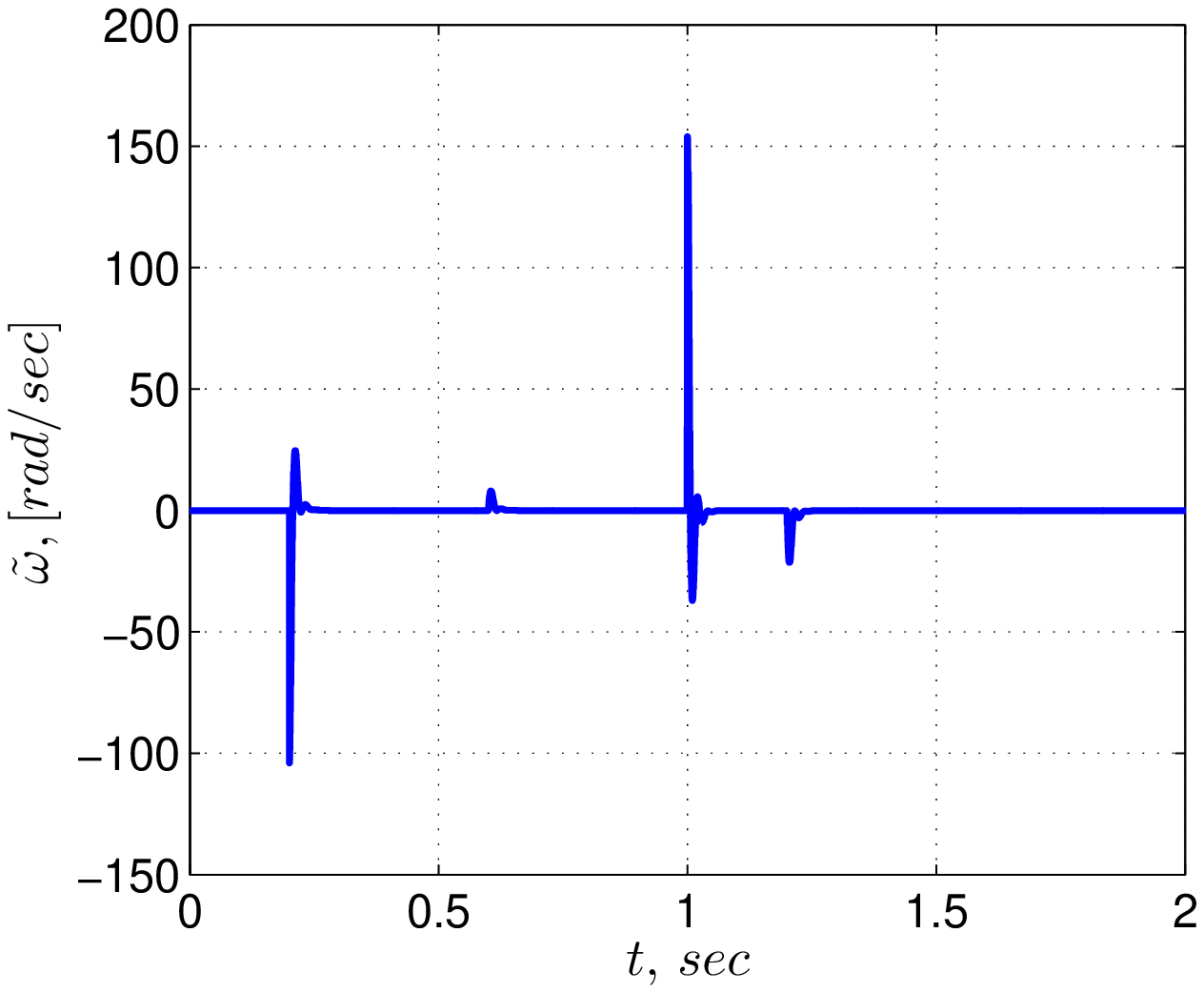}}
\
\subfloat[][Current errors $\tilde{i}_d$ (I) and $\tilde{i}_q$ (II)]{{\label{fig_12b}}\includegraphics[width=0.31\textwidth]{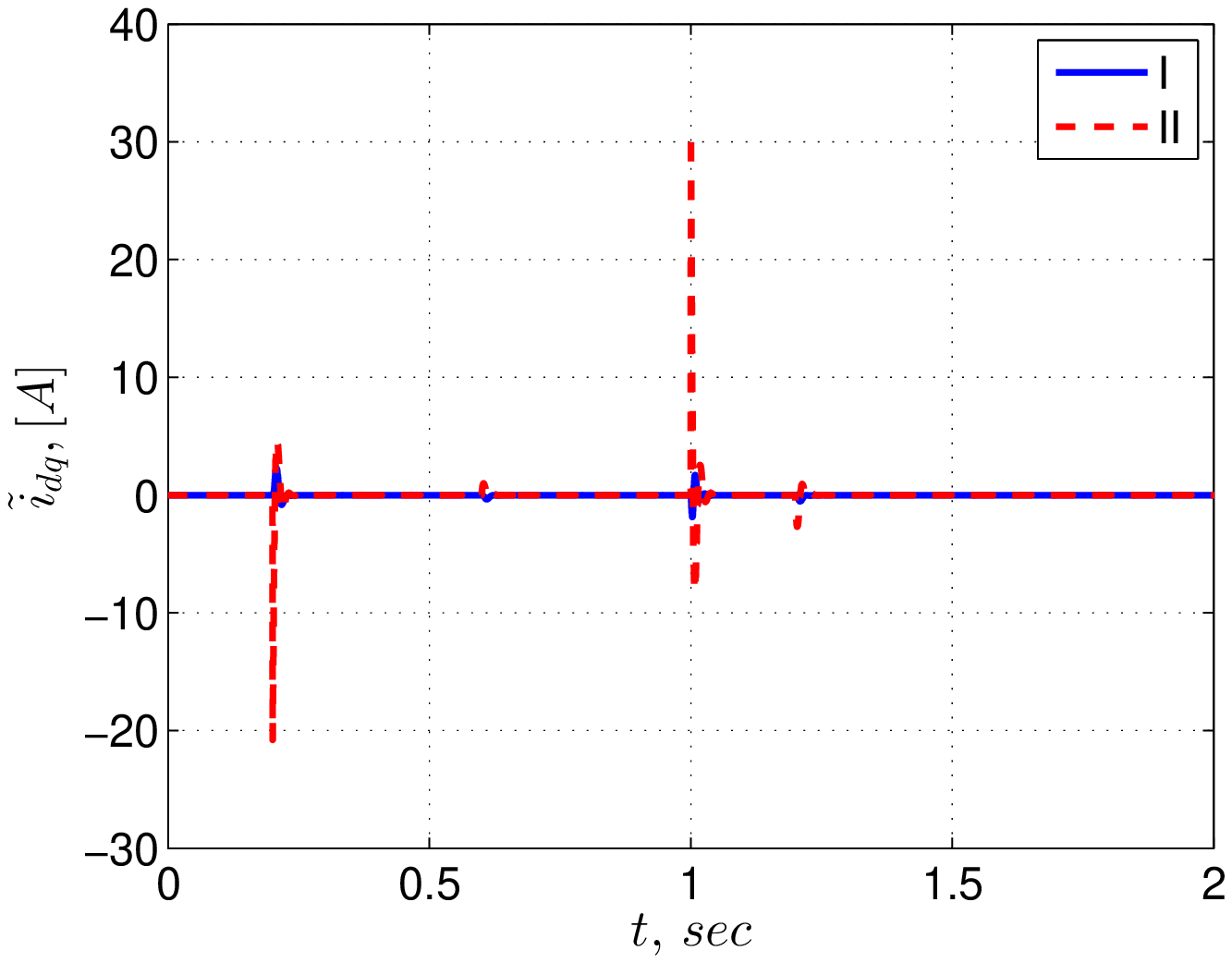}}
\
\subfloat[][Voltages $v_d$ (I) and $v_q$ (II)]{{\label{fig_12c}}\includegraphics[width=0.31\textwidth]{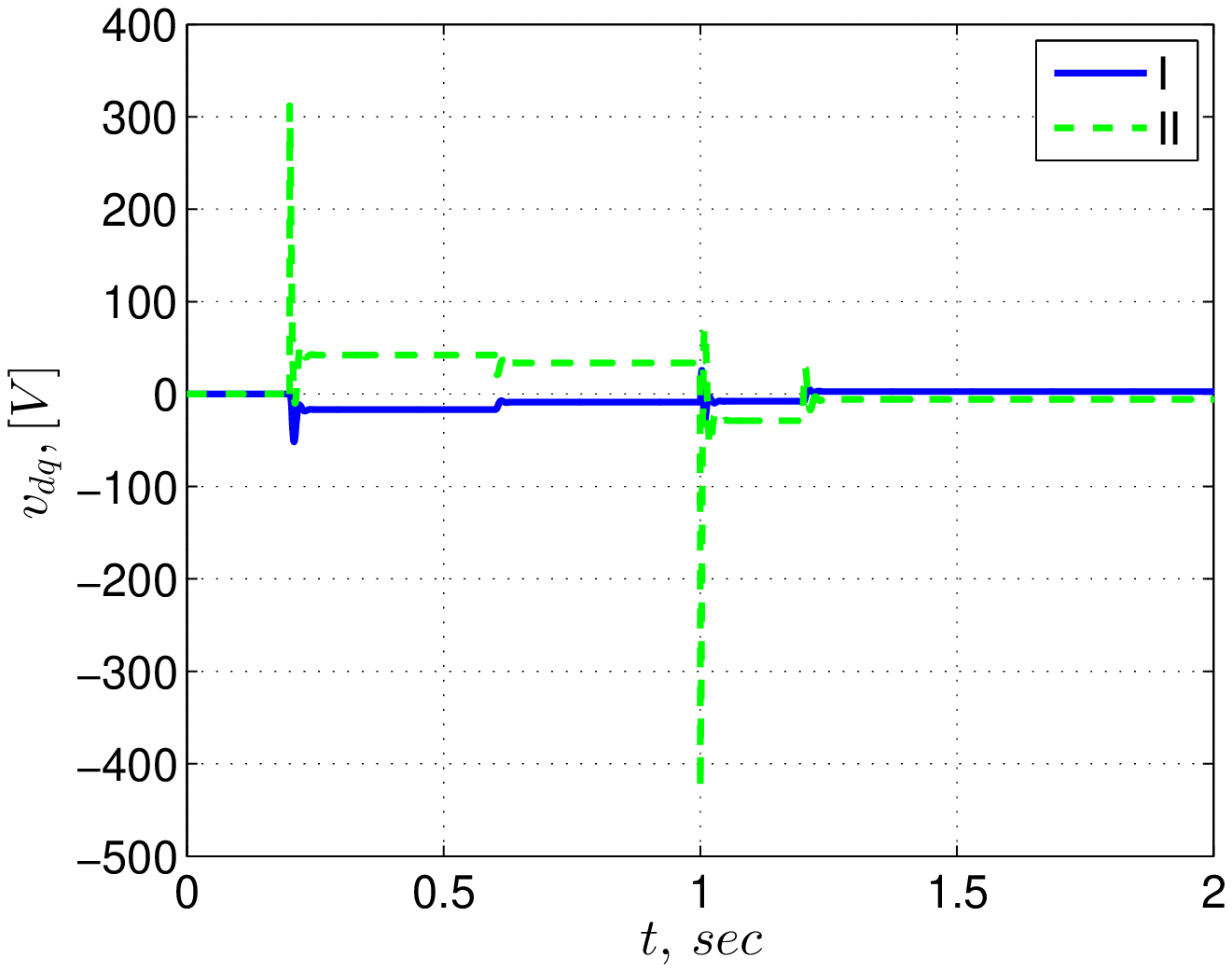}}
\vspace{-2mm}
\caption{Transients in the system with inner-loop current PI \eqref{pi} and outer-loop speed PI \eqref{nespi},  load torque of Fig. \ref{fig_2} and speed reference of Fig. \ref{fig_1}: $k_p=15$, $k_i=2000$, $a_p=0.2$ and $a_i=18$.}
\label{fig_12}
\end{figure*}

Fig. \ref{fig_12} shows the deleterious effect of increasing the gains $a_p$ and $a_i$ of the outer-loop PI. Indeed, although this results in a better transient behaviour of the speed error, it yields unrealistic overshoots both in motor currents and voltages, shown in Figs. \ref{fig_12b} and \ref{fig_12c}, correspondingly.

\section{Conclusions and Future Research}
\lab{sec7}
%
We have established the practically interesting---though not surprising---result that the PMSM can be globally regulated around a desired equilibrium point with a simple (adaptive) PI control around the current errors, provided some viscous friction is present in the rotor dynamics and the proportional gain of the PI is suitably chosen. The key ingredient to establish this result is the proof in Lemma \ref{lem1} that the incremental model of the PMSM satisfies the dissipation inequality \eqref{disine}. Our main results are established with simple calculations and invoking elementary Lyapunov theory with the natural---quadratic in the increments---Lyapunov functions.

Some topics of current research are the following.

\noindent - From the theoretical viewpoint the main drawback of the results reported in the paper are the requirement of existence, and knowledge, of the friction coefficient $R_m$. As shown in Lemma \ref{lem3} the requirement of knowing $R_m$ can be relaxed---at the price of complicating the controller and requiring some excitation conditions. However the assumption of $R_m>0$ seems unavoidable if we want to preserve a a simple PI structure, see Remark \ref{rem8}. It should be underscored, however, that from the practical viewpoint, the assumption that the mechanical dynamics has some static friction---that may be arbitrarily small---is far from being unreasonable.

\noindent - As discussed in \cite{ZONORTBEN} in the context of power systems, the absence of the outer-loop PI significantly deteriorates the transient performance of the inner-loop PI. A similar situation appears here for the PMSM.\footnote{The authors thank the anonymous Reviewer $\#$3 for bringing this issue to our attention.}. Unfortunately, the analysis of the classical outer-loop PI in speed errors \eqref{nespi} is hampered by the lack of a convergence proof of the estimation error.
 
\noindent - In the case of $L_d \neq L_q$ torque can be made even larger by an additional reluctance component $x_1^\star   \neq 0$. The implications of this choice on the passivity of the incremental model remains to be investigated.

\noindent - The extension of the result to the case of salient PMSM is also very challenging---see \cite{ICHetal} for the corresponding $\alpha\beta$ model.

\noindent - The lower bound on the proportional gain can be computed invoking the physically reasonable Assumption \ref{ass1}. However, the reference value for $i_q$ is dependent on $\tau_L$. As shown in Proposition \ref{pro2} this problem can be solved using an adaptive PI, at the high cost of  knowledge of the PMSM model parameters.

\noindent - Experimental results of PI current control abound in the literature and experiments of an observer, similar to \eqref{tauobs}, may be found in \cite{PETetal}. However, it would be interesting to validate experimentally the performance of the proposed adaptive PI and, in particular, investigate how it compares with the classical outer-loop speed PI \eqref{nespi}.    
%

\end{document}